\documentclass{aa}  

\usepackage{graphicx}
\usepackage{txfonts}

\usepackage{epstopdf}
\usepackage{amsmath}
\usepackage[usenames, dvipsnames]{xcolor}
\usepackage{textcomp,gensymb}
\usepackage[normalem]{ulem}
\usepackage{natbib}
\usepackage{amsmath}
\usepackage{hyperref}
\usepackage{listings}
\usepackage{color}

\definecolor{dkgreen}{rgb}{0,0.6,0}
\definecolor{gray}{rgb}{0.5,0.5,0.5}
\definecolor{mauve}{rgb}{0.58,0,0.82}
\definecolor{golden}{rgb}{0.86,0.65,0.01}

\defcitealias{Antoja2020}{A20}
\def\kms{{\rm km\,s^{-1}}}
\def\masyr{{\rm mas\,yr^{-1}}}

\def\HP{HEALpix }
\def\HPf{HEALpix}

\def\Gaia{{\it Gaia }}
\def\Gaiaf{{\it Gaia}}

\begin{document}

   \title{The outer disc in shambles: blind detection of Monoceros and ACS with Gaia's astrometric sample}

   \titlerunning{Monoceros, ACS and other outer disc structures with \Gaia astrometry}

   \author{P. Ramos\inst{1,2}\fnmsep\thanks{email: p.ramos@unistra.fr}
            \and
          T. Antoja\inst{1}
            \and
          C. Mateu\inst{3}
                \and
          F. Anders\inst{1}
                \and
          C. F. P. Laporte\inst{4}
                \and
        J.A. Carballo-Bello\inst{5}
                \and
         B. Famaey\inst{2}
                \and
        R. Ibata\inst{2}
          }

   \institute{Dept. FQA, Institut de Ci{\`e}ncies del Cosmos (ICCUB), Universitat de Barcelona (IEEC-UB), Mart{\'i} Franqu{\`e}s 1, E-08028 Barcelona, Spain
        \and
            Observatoire astronomique de Strasbourg, Universit{\'e} de Strasbourg, CNRS, 11 rue de l’Universit{\'e}, 67000 Strasbourg, France
         \and
             Departamento de Astronom{\'i}a, Instituto de F{\'i}sica, Universidad de la Rep{\'u}blica, Igu{\'a} 4225, CP 11400 Montevideo, Uruguay
        \and
             Kavli Institute for the Physics and Mathematics of the Universe (WPI), The University of Tokyo Institutes for Advanced Study (UTIAS), The University of Tokyo, Chiba 277-8583, Japan
        \and
            Instituto de Alta Investigaci{\'o}n, Universidad de Tarapac{\'a}, Casilla 7D, Arica, Chile
            }
             
   \date{Received XXX; accepted XXX}

 
  \abstract
   {The astrometric sample of \Gaia allows us to study the  outermost Galactic disc, the halo and their interface. It is precisely at the very edge of the disc where the effects of external perturbations are expected to be the most noticeable.}
   {Our goal is to detect the kinematic substructure present in the halo and at the edge of the Milky Way (MW) disc, and provide observational constraints on their phase-space distribution.}
   {We download, one \HP at a time, the proper motion histogram of distant stars, to which we apply a Wavelet Transformation to reveal the significant overdensities. We then analyse the large coherent structures that appear in the sky.}
   {We reveal a sharp yet complex anticentre dominated by Monoceros (MNC) and the Anticentre Stream (ACS) in the north, which we find with an intensity comparable to the Magellanic clouds and the Sagittarius stream, and by MNC south and TriAnd at negative latitudes. Our method allows us to perform a morphological analysis of MNC and ACS, both spanning more than 100$^\circ$ in longitude, and to provide a high purity sample of giants with which we track MNC down to latitudes as low as $\sim$5$^\circ$. Their colour-magnitude diagram is consistent with extended structures at a distance of $\sim$10-11 kpc originated in the disc, with a very low ratio of RR Lyrae over M giants, and kinematics compatible with the rotation curve at those distances or only slightly slower.}
   {We present a precise characterisation of MNC and ACS, two previously known structures that our method reveals naturally, allowing us to detect them without limiting ourselves to a particular stellar type and, for the first time, using only kinematics. Our results allow future studies to model their chemo-dynamics and evolution, thus constraining some of the most influential processes that shaped the MW.}


   \maketitle
%

\section{Introduction}\label{sec:healpix_intro}

Most of the studies that discovered new substructures within the second data release \citep[DR2,][]{dr2} of the \Gaia mission \citep{gaiamission} used the full 6D phase-space sample, and their impact on our current understanding of the Milky Way (MW) and its history is undeniable. Good examples of that are the work of \citet{Belokurov2018}, \citet{GaiaBabusiaux2018}, \citet{Haywood2018} and \citet{Helmi2018}, which identified a large group of stars accreted in the last major merger event of the MW that took place $\sim$10\,Gyr ago. Another example is the advance in the study of the \emph{moving groups} and the possibility to now visualise the kinematic substructure directly in the plane of Galactocentric radii against rotational velocity with the ridges \citep{Antoja2018,Kawata2018,Ramos2018,Laporte2019b,Fragkoudi2019,Khanna2019}. Nevertheless, this sample is limited to $G$<\,13\,mag approximately (above that, the completeness drops significantly), restricting the explotation of the kinematic data to a volume of $\sim$3\,kpc radius from the Sun. Despite some attempts to extend the kinematic maps to further distances either by using statistical corrections to the parallax \citep{Lopez-Corredoira2019,Lopez-Corredoira2020}, or by adding photometric \citep{Anders2019} or spectroscopic information \citep{Liu2017,Wang2019}, these only have a significant amount of stars up to Galactocentric radii of $\sim$16\,kpc.

In contrast, the 5D sample (only astrometry and no radial velocity) is more than two orders of magnitude larger than the 6D one. Its power is exemplified by the work of, for instance, \citet{Castro-Ginard2018,Castro-Ginard2020} by discovering hundreds of new open clusters throughout the disc, \citet{Malhan2018c} and \citet{Ibata2019} by revealing several new tidal streams in the halo, or by the large sample of halo stars selected using a combination of photometry and proper motions provided in \citet{Koppelman2020}. Another good example is the detection of the Sagittarius \citep[Sgr,][]{Ibata1994} stream using mostly its kinematic signature \citep{Antoja2020,Ibata2020,Ramos2020}.

The astrometric sample reaches down to $G\sim$21\,mag, expanding the volume probed significantly\footnote{Using red clump stars, and assuming no extinction, we can potentially reach up to distances of roughly 100\,kpc from the Sun.}, meaning that we can use it to trace kinematic structures well into the halo. Among the different stellar systems that we expect to find within the 5D sample we count globular clusters \citep[e.g.,][]{Baumgardt2019}, streams \citep[e.g.,][]{Belokurov2006b}, dwarf galaxies and even ultra faint dwarf galaxies \citep[e.g.,][]{Willman2005a,Belokurov2007,Koposov2015}. Moreover, this sample also covers the outermost regions of the MW disc, where \citet{Newberg2002} reported, almost two decades ago, the presence of a peculiar population above the mid-plane of the Galaxy, bluer than the thick disc and clearly appreciable as an overdensity of Main Sequence Turn-off (MSTO) stars at a distance of $\sim$10\,kpc from the Sun. Known as Monoceros (MNC, also referred to as Galactic anticentre stellar structure or GASS), this structure was observed to span more than a hundred degrees in longitude \citep[100$^\circ$ < l < 270$^\circ$, see, e.g.,][]{Rocha-Pinto2004,Morganson2016} both in the north and south hemispheres. During the past two decades, there has been an intense debate over its origin, in part due to the difficulties of confronting the data with the different models available \citep{Slater2014}. Out of the many possible mechanisms proposed by \citet{Ibata2003}, there have been mainly two leading hypotheses: accretion and disc perturbation. 

The idea that MNC is the tidal debris of an accreted satellite was based on its morphology (it looks like a stream) and on its metallicity and kinematics \citep[e.g.,][]{Yanny2003,Crane2003,Wilhelm2005,Conn2005}, partially supported by the simulations of \citet{Helmi2003} and \citet{Penarrubia2005}. This lead to the hunt for its progenitor and, after discarding the Canis Major \citep{Martin2004} over-density as a candidate \citep[e.g.,][]{Momany2006,Rocha-Pinto2006,Carballo-Bello2020}, none has yet been found, despite the attempts to detect the continuation of the hypothetical tidal stream at other Galactic latitudes \citep{Conn2007,Conn2008}. Nevertheless, upper limits for the total mass of the progenitor have been calculated \citep[e.g.,][]{Guglielmo2018}. 

On the other hand, several works have shown that the close passage of a satellite can induce significant substructure in the outer parts of the disc \citep[e.g.,][]{Younger2008,Purcell2011,Gomez2016}. The interaction with a dwarf galaxy as massive as Sgr could cause some of the disc material to move to more inclined and eccentric orbits, and produce a stream of stars consistent with the observations \citep[see also][ where, instead of a single satellite, the perturbers are 6 dark matter subhalos of masses $\sim$10$^{10}$ M$_\odot$]{Kazantzidis2008}. 
More recently, the simulations by \citet{Laporte2018,Laporte2019} have shown that it is possible to create extended structures similar to MNC as well as feather-like structures during a satellite encounter, while at the same time reproducing qualitatively part of the phase-space substructure observed at the solar neighbourhood. The detection of a vertical wave-like pattern in the disc \citep{Widrow2012} that propagates almost radially \citep{Xu2015,Schonrich2018}, in agreement with the simulations of \citet{Gomez2013}, and the discovery of the phase-space spiral and consequent confirmation that our Galaxy is undergoing phase-mixing \citep{Antoja2018} further supports the perturbative scenario.

Other structures in the outer disc are the Anticentre Stream (ACS) and Eastern Band Structure \citep[EBS,][]{Grillmair2006}, or the Triangulum-Andromeda (TriAnd1 and TriAnd2) overdensities \citep{Majewski2004,Rocha-Pinto2004,Martin2007}. The connection between all of these and MNC has also been subject to scrutiny for many years and is still not entirely clear \citep[but see the models of][]{Xia2015,Sheffield2018}. For instance EBS, which was described as an independent structure by \citet{Grillmair2011}, has now been suggested to be part of the MNC ring by \citet{Deason2018} using a combination of \Gaia and SDSS \citep{York2000} data. \citet{deBoer2018} re-analysed the kinematics of MNC and ACS with SDSS astrometry calibrated with \Gaia DR1 to provide accurate kinematic maps, showing that they have similar yet clearly distinct kinematic trends that can be used to establish the processes that form them. Very recently, \citet{Laporte2020} studied the [Mg/Fe]-[Fe/H] distribution of ACS and MNC with a combination of \Gaia DR2 data and LAMOST-SEGUE-APOGEE to, guided also by their previous simulations of an isolated MW interacting with Sgr, intercede in favour of a disc origin for the two structures. In their work, they also show with colour-magnitude diagrams (CMDs) that both have a conspicuous red clump (RC), in contrast with previous studies that mainly focused on the Main Sequence (MS), the MSTO or the 2MASS M-giants. Here we aim to provide an independent detection and characterisation of these structures that can help us clarify their true extent and 3D morphology, as well as their nature. A deeper understanding of the events that lead to the observed stellar distribution in the anticentre could be used to constrain the orbit and mass of Sgr, as well as its effect on the gas and stars of our Galaxy.

In this work, we search for substructure following the strategy devised by \citet{Antoja2015b}. The original goal of this method was to detect ultra faint dwarf galaxies in the halo using the fact that these should create, simultaneously, an overdensity in proper motion space and in the sky. Here we use the first half of the methodology, that is, its application in proper motion space only, to find the kinematic substructure at large heliocentric distances. This approach allows us to scan the whole Celestial sphere systematically, homogeneously, and using a statically robust technique that can distinguish small but significant overdensities in proper motion space, and track their changes as we move with Galactic longitude and latitude. As a result, our all-sky maps are dominated by three large structures: the Magellanic clouds, the Sgr stream \citep[as reported in][]{Antoja2020}, and MNC-ACS. 
Our goal is to map and study the morphology and kinematics of the structures in the Galactic anticentre, taking advantage of our methodology which allows us detect them and obtain a large set of members with (almost) no prior information. 

This paper is organised as follows. In Sect.~\ref{sec:data&methods} we describe the strategy used to process the large amount of data available with \Gaiaf. Section~\ref{sec:healpix_results} then enumerates the different systems detected with our method and, in Sect.~\ref{sec:acs}, we focus on characterising the complex kinematics of the anticentre, specially in the north where we observe MNC and ACS. We discuss the implications of our findings in Sect.~\ref{sec:healpix_disc}, and finally present our conclusions in Sect.~\ref{sec:healpix_conc}.

\section{Data and methods}\label{sec:data&methods}

\begin{figure*}[]
   \centering
    \includegraphics[width=0.98\textwidth]{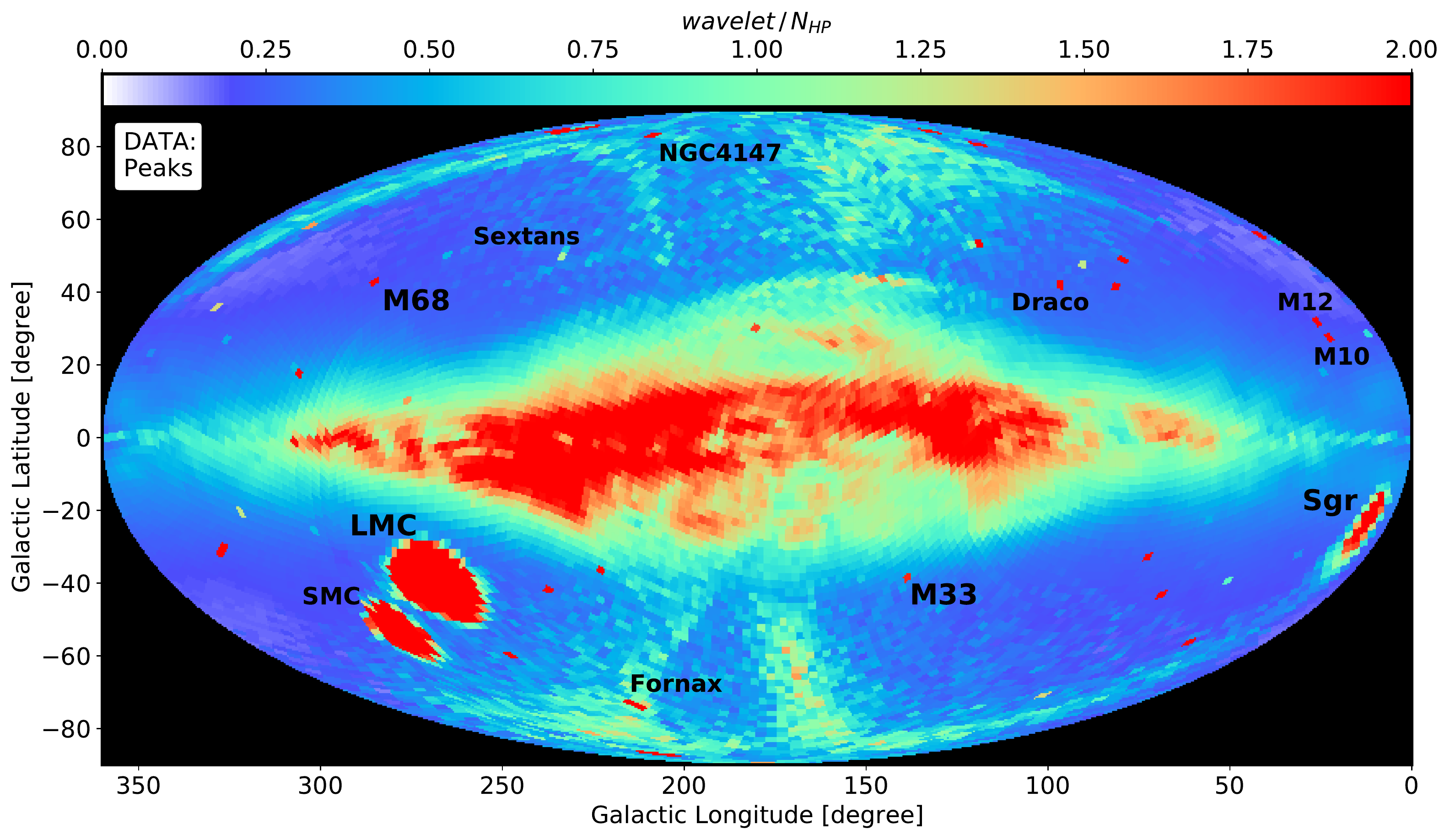}
    \caption{Mollweide projection of the relative intensity of the dominant structure in the proper motion plane at each \HPf. By showing only the most significant kinematic overdensity and normalising to the number of stars in the \HPf, a large number of structures become visible: the Sgr stream, tens of globular clusters and an intricate anticentre. We have labelled some of most relevant ones.}
    \label{fig:allsky-wt} 
\end{figure*}

In this work we use the same sample and methodology presented in \citet[hereafter A20]{Antoja2020}, which we reproduce here for convenience. We exploit the full \Gaia DR2 \citep{dr2}, not restricting ourselves to any magnitude limit other than the one intrinsic to the instruments, and applying only the following two filters:

\begin{equation}\label{eq:parallax_filter}
    \varpi - \sigma_\varpi < 0.1\,\text{mas, } G_{BP}-G_{RP} > 0.2\,\text{mag,}
\end{equation}

\noindent aimed at reducing the level of foreground contamination, that is, nearby stars that block our view of the halo and outer disc.

The resulting sample contains 700\,412\,152 sources. Properly processing and analysing such a large data set is obviously impractical to do with a regular desktop computer. Usually, it would require a Big Data infrastructure. Nevertheless, given that we want to study changes in the velocity planes, our observables are the proper motion histograms themselves. Hence, we download them in parallel directly from the \Gaia Archive\footnote{Hosted at: \href{https://gea.esac.esa.int/archive/}{https://gea.esac.esa.int/archive/}}, with the query  

\begin{lstlisting}
SELECT COUNT(*) as N, pmra_index*BINSIZE as pmra, pmdec_index*BINSIZE as pmdec FROM (SELECT source_id, FLOOR(pmra/BINSIZE) AS pmra_index, FLOOR(pmdec/BINSIZE) AS pmdec_index FROM gaiadr2.gaia_source WHERE source_id BETWEEN  HPNUM*2**35*4**(12-LVL)  AND  (HPNUM+1)*2**35*4**(12-LVL) AND parallax-parallax_error < 0.1 AND bp_rp >= 0.2 AND pmra IS NOT NULL AND pmdec IS NOT NULL) as sub GROUP BY pmra_index, pmdec_index
\end{lstlisting}

\noindent where \emph{LVL} is the level of the \HP grid (here, 5), \emph{HPNUM} is the \HP to be processed, and \emph{BINSIZE} is the size of the histogram binning (here 0.24\,$\masyr$).

Then, we apply the Wavelet Transformation \citep[WT,][]{Starck2002}, followed by a peak detection algorithm, to each of the 12\,288 histograms we have downloaded to detect the significant kinematic structures (assuming Poisson noise). The resulting wavelet coefficient of each peak is then, by construction, proportional to its density in proper motion space. To simplify the analysis, we only keep one structure at each \HPf, the one with the highest relative intensity ($WT$/$N_{hp}$):

\begin{equation}\label{RI}
    \frac{WT}{N_{\mathrm{hp}}}\times 1000,
\end{equation}
\noindent where $N_{\mathrm{hp}}$ is the total number of sources in the \HP and is used to normalise the wavelet coefficient \citepalias[for more details on the method, see][]{Antoja2020}.

\section{Global map of the substructures}\label{sec:healpix_results}

Figure~\ref{fig:allsky-wt} shows the Mollweide projection of the sky in Galactic coordinates coloured by the relative intensity (Eq.~\ref{RI}) of the highest peak in the proper motion histogram. By selecting only the overdensity with the largest intensity present at each proper motion histogram, we can focus on the dominant kinematic structure of the \HPf\footnote{In some cases, the dominant peak might not be the peak in the proper motion plane with the largest amount of stars inside it.}. The normalisation used in Eq.~\ref{RI} compensates the density gradient of the Galaxy and gives more contrast to the structures at higher latitudes. This figure reveals a wealth of substructure that cannot be seen with a simple density map of our sample. For instance, Fig.~\ref{fig:allsky} contains the number of sources that pass our filters at each \HP (top panel) where we can only identify the Magellanic clouds, some globular clusters and the imprints of the extinction (which is shown in the middle panel for comparison, \citealt{Schlegel1998}) or the \Gaiaf\  scanning law (bottom panel). 

With Fig.~\ref{fig:allsky-wt} we have been able to detect:

\begin{itemize}
    \item The Magellanic clouds: Their angular size in the kinematic maps, in contrast with their apparent angular size in the star counts map of Fig.~\ref{fig:allsky} (top panel), is larger and shows the true extent of these systems as already noted in, e.g. \citet{Helmi2018b}. We also detect substructure within them (not shown here), and we are able to recover some of the globular clusters\footnote{More specifically, we have detected \emph{Magellanic Halos’s Clusters} as they are described in \citet{Bica2019}.} that orbit the Large Magellanic cloud, like the recently detected Gaia 3 \citep{Torrealba2019}.
    
    \item The Sgr stream: The core of this dwarf galaxy and its stream are also clearly visible in our map. Interestingly enough, we do not observe it in the top panel of Fig.~\ref{fig:allsky}, which highlights the difficulty, even in the \Gaia era, to detect this structure with just stellar counts. 
    
    \item Almost vertically mirrored to the Sgr stream, we note a feature with a stream-like shape.  The CMDs of the sources that produce these peaks do not present any coherent isochrone-like shape. Instead, they appear clumped around the faint and blue corner of the diagram. Added to the fact that their proper motions are always normally distributed around the origin, regardless of the position in the sky, this leads us to conclude that these sources are actually quasars. Although quasars are ubiquitous and should not produce a band in the sky, the scanning law of \Gaia favours certain regions of the sky as can be seen in the bottom panel of Fig.~\ref{fig:allsky} with the {\normalfont\ttfamily{astrometric\_gof\_al}}\footnote{This 'gaussianized chi-square' is an indicator of the quality of the astrometric solution. Values above +3, thus, indicate a bad fit to the data. Other indicators are {\normalfont\ttfamily{astrometric\_excess\_noise}} or {\normalfont\ttfamily{astrometric\_n\_good\_obs\_al}}.} that quantifies the quality of the astrometric solution. In these parts, the astrometric uncertainties of the  quasars are low enough so that they become more relevant than the diffuse halo population of stars. Since we did not expect to obtain such a clear signal from the  quasars, we did not remove them beforehand. Nevertheless, our results are not affected by their presence. With the list of extended objects that will be published in \Gaia DR3, we will be able to remove these objects up-front already within the queries.
    
\begin{figure}
   \centering
      \includegraphics[width=0.5\textwidth]{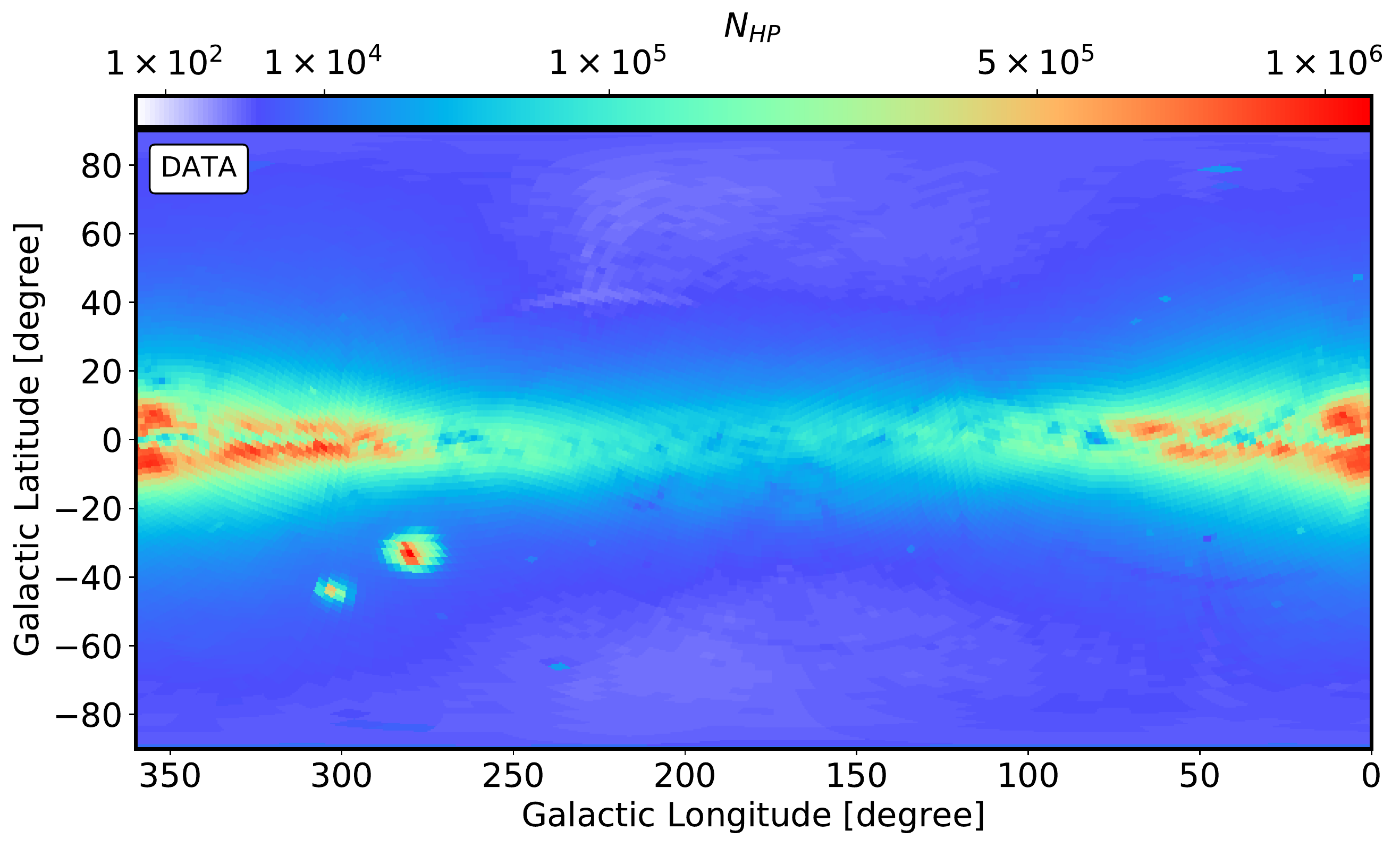}
      \includegraphics[width=0.5\textwidth]{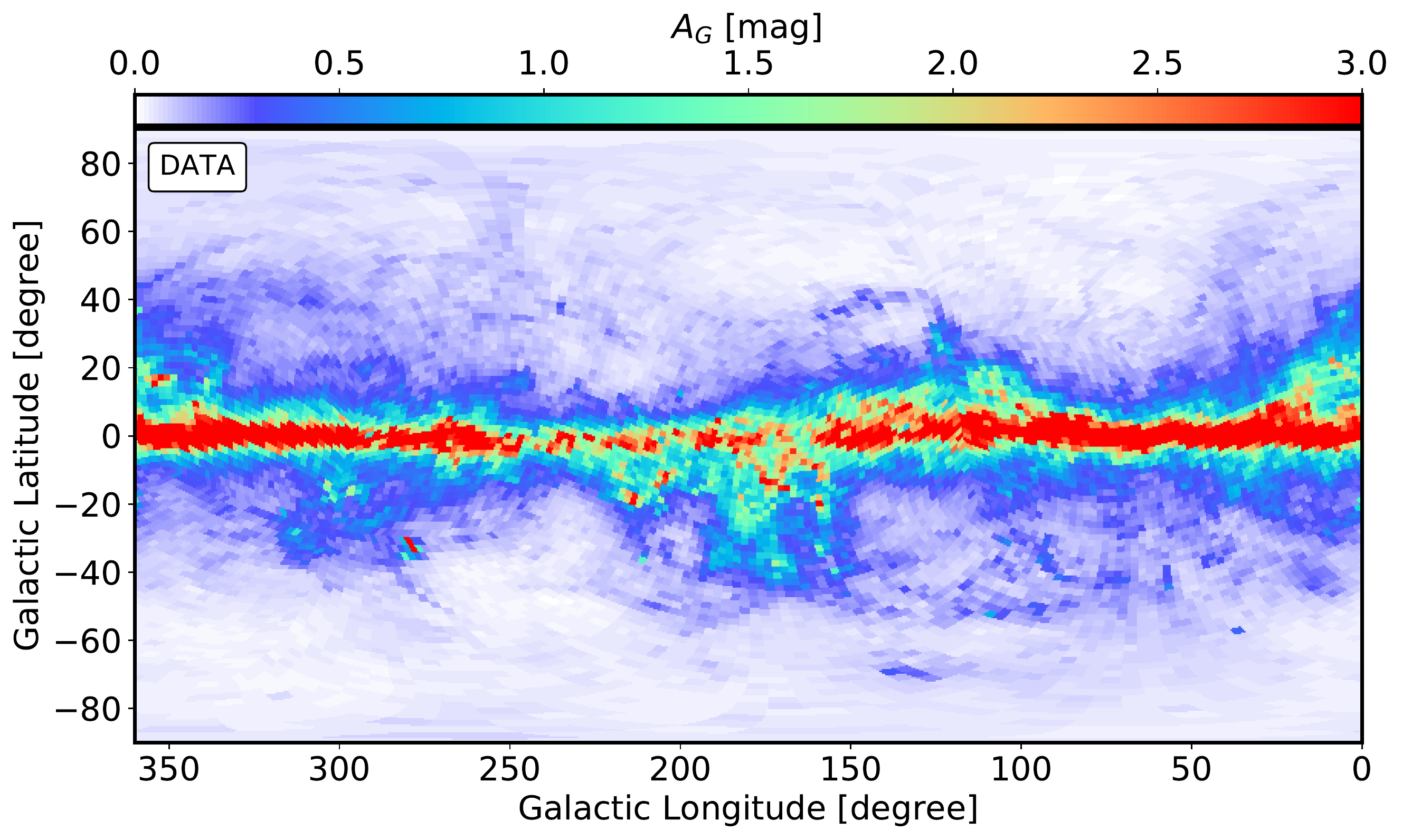}
      \includegraphics[width=0.5\textwidth]{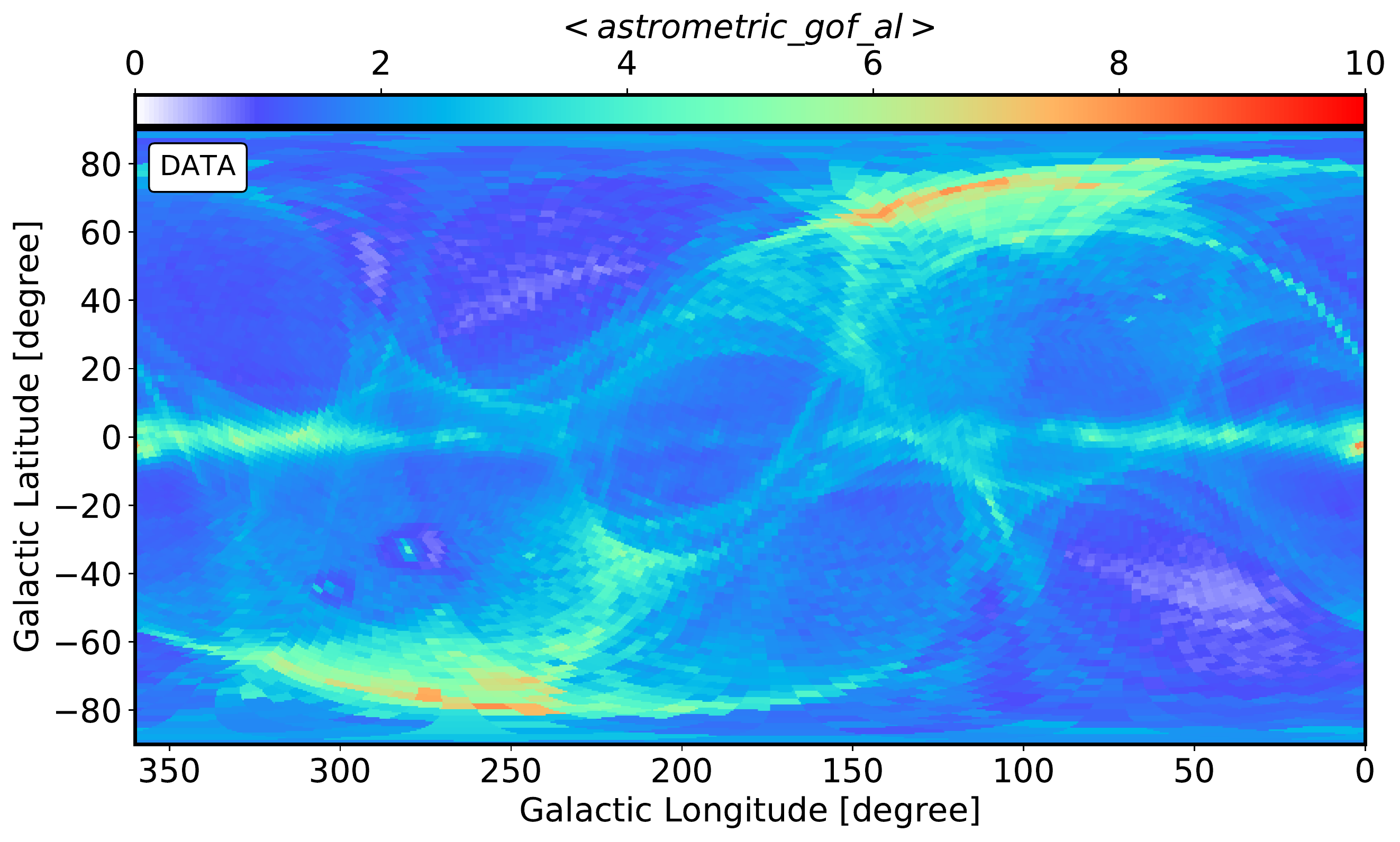}
      \caption{Top: Number of stars in each \HP that fulfil the selection described in Sect.~\ref{sec:data&methods}. Middle: Absorption at infinite in the $G$ band  at each \HPf, obtained from the \citet{Schlegel1998} maps with the re-calibration by \citet{Schlafly2011} and using the mean $G_{BP}-G_{RP}$ colour in the \HP together with the transformations described in Appendix~A of \citet{Ramos2020}. Bottom: Average {\normalfont\ttfamily{astrometric\_gof\_al}} at each \HP (see Appendix~\ref{app:queries}).}
      \label{fig:allsky} 
\end{figure}

\begin{figure}
   \centering
    \includegraphics[width=0.5\textwidth]{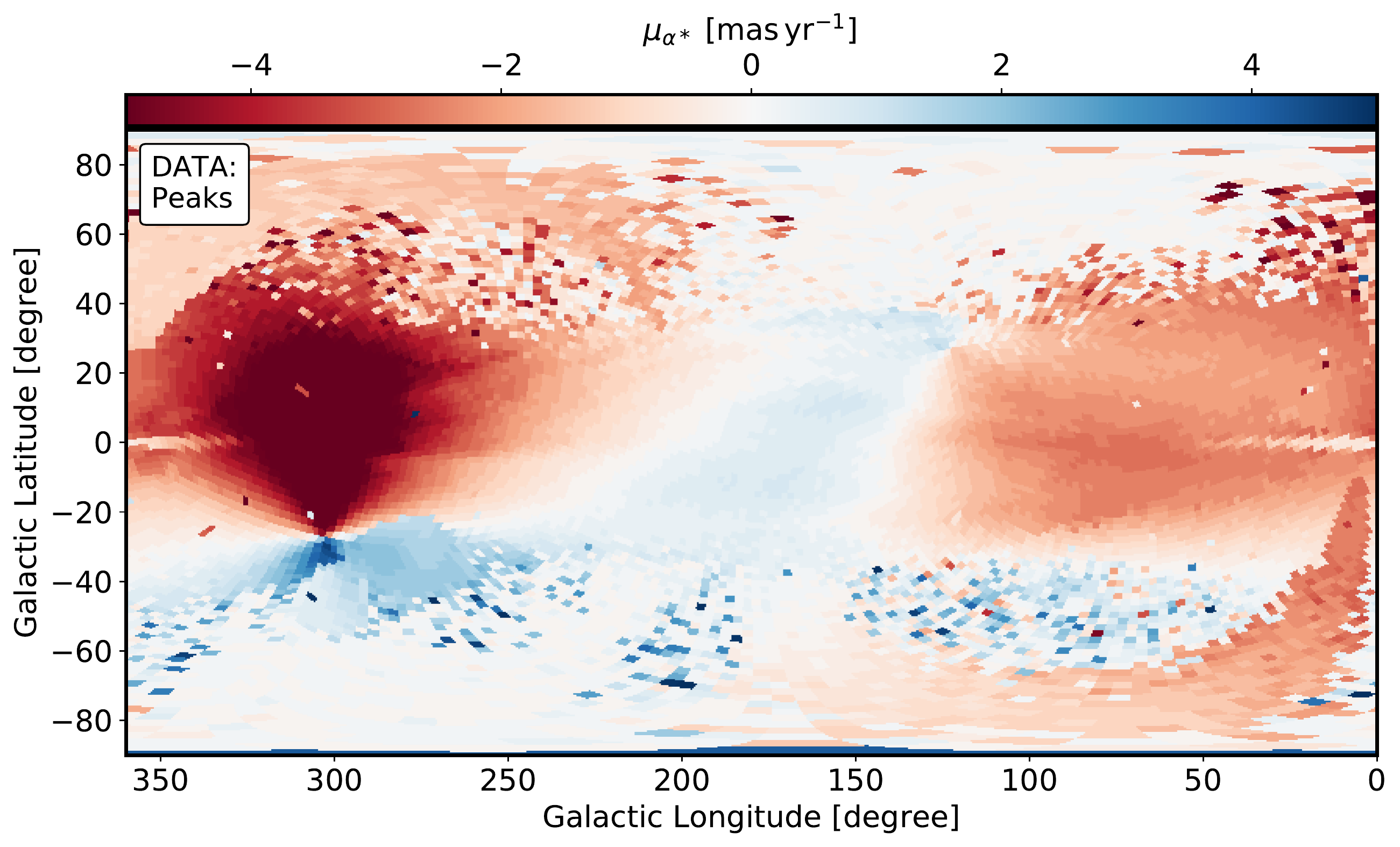}
    \includegraphics[width=0.5\textwidth]{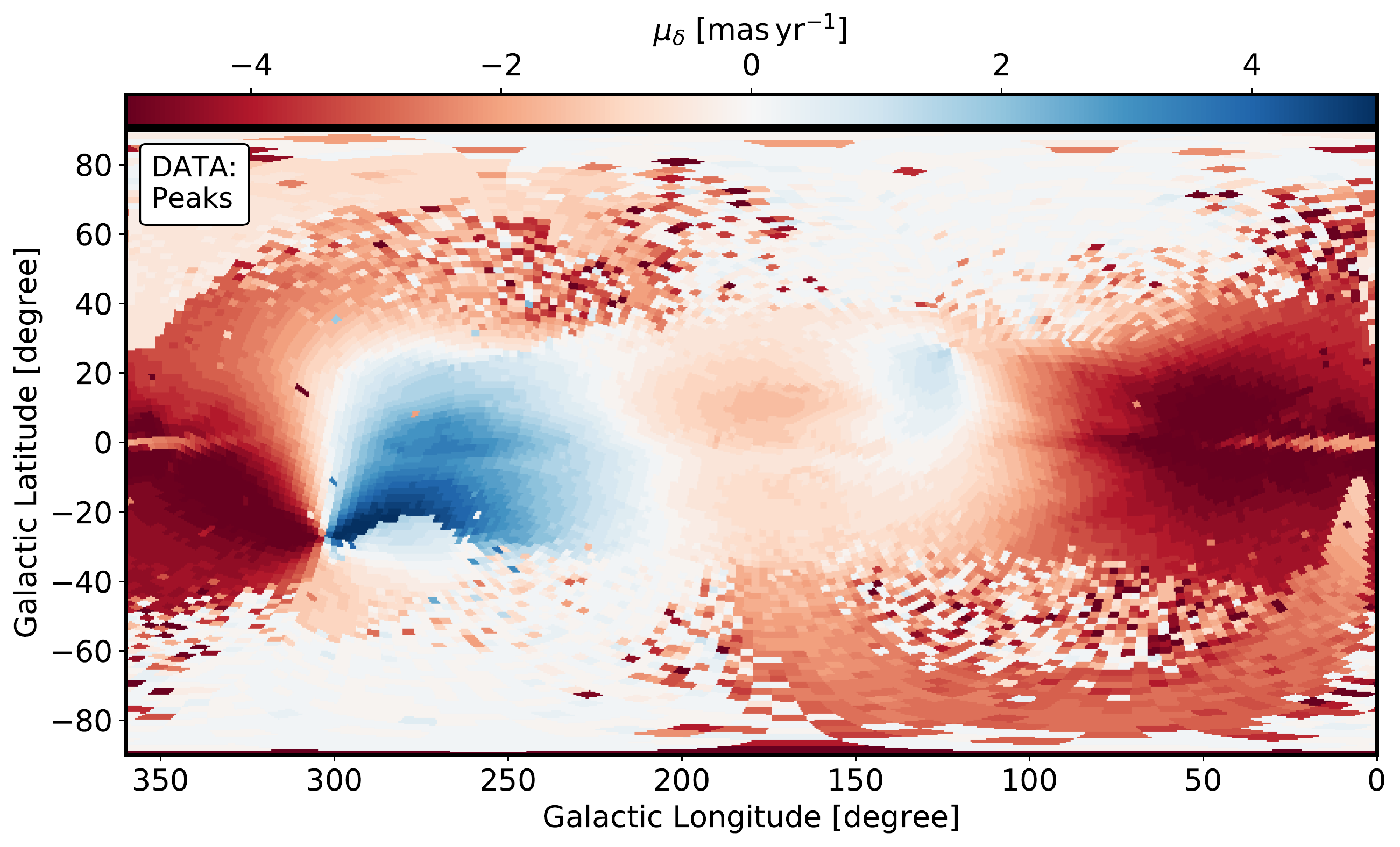}
    \caption{Proper motion coordinates of the dominant peak at each \HPf. Top: Proper motion in right ascension. Bottom: Same but in declination. Conspicuous stream-like patterns crossing the entire Celestial sphere can be appreciated, one of which is Sgr (bottom right to top left) and the other corresponds to the  quasars (bottom left to top right).}
    \label{fig:allsky-pm} 
\end{figure}
    
    \item Nearby galaxies: Apart from the Magellanic clouds, we also detect M31 and M33, the latter clearly visible in Fig.~\ref{fig:allsky-wt} at (l,\,b) $\sim$ (133$^\circ$,\,-31$^\circ$).
    
    \item Dwarf spheroidals: Our method is able to detect several dwarf spheroidals, like Fornax, Sextans or Sculptor, as well as fainter ones like Draco. 
    
    \item Globular clusters: We recover 51 globular clusters from the \citet{Bica2019} catalogue. From their 200 objects classified as globular clusters, more than half are in the bulge where we do not detect any either because the contrast with the foreground is too low or due to the cut in parallax applied.
    
    \item Ultra Faint Dwarf galaxies: We do not recover any of the known Ultra Faint Dwarfs galaxies based solely on their kinematic signature. Their proper motion uncertainties are too large and there are too few members \citep[see][]{Massari2018} to produce a significant overdensity. Nevertheless, we note that when we apply the full methodology described in \citet{Antoja2015b}, that includes the search of peaks in the sky and not only in proper motion as done here, we can effectively recover most of them.
    
    \item Anti-centre: Apart from all the substructure we find in the halo, our methodology reveals complex kinematic substructures towards the anticentre of the MW, dominated by two arch-like features in the north Galactic hemisphere. After comparing with the extinction map shown in the middle panel of Fig.~\ref{fig:allsky}, we confirm that these features are not aligned with regions of high absorption. Also, we have checked how the {\normalfont\ttfamily{astrometric\_gof\_al}} map (bottom panel of Fig.~\ref{fig:allsky}) superposes to the WT intensity map, from which we conclude that the shape of the bottom arch is artificially enhanced by the scanning law. The cavity at $\ell\sim$180$^\circ$, $b\sim$~20$^\circ$ coincides with a region poorly sampled by \Gaia and therefore the intensity (proportional to stellar counts) is lower.

\end{itemize}

We note that some of these structures also appear in Fig.~\ref{fig:allsky-pm}, where we colour the sky according to the proper motion of the highest peak (the peak used to colour Fig.~\ref{fig:allsky-wt}). The most clear one is the Sgr stream, which we have analysed in detail in \citetalias{Antoja2020}. Also, the structure at latitude b$\sim$35$^\circ$ (140$^\circ$\,<\,l\,<\,200$^\circ$) appears as a conspicuous arch in the proper motion map. We devote the following sections to the analysis and characterisation of the kinematic substructure present at the outer disc, focusing mostly in the north where we observe these two conspicuous arches already mentioned. 

\section{Kinematic features in the anticentre}\label{sec:acs}

\begin{figure*}[t]
    \centering
    \includegraphics[width=0.7\textwidth]{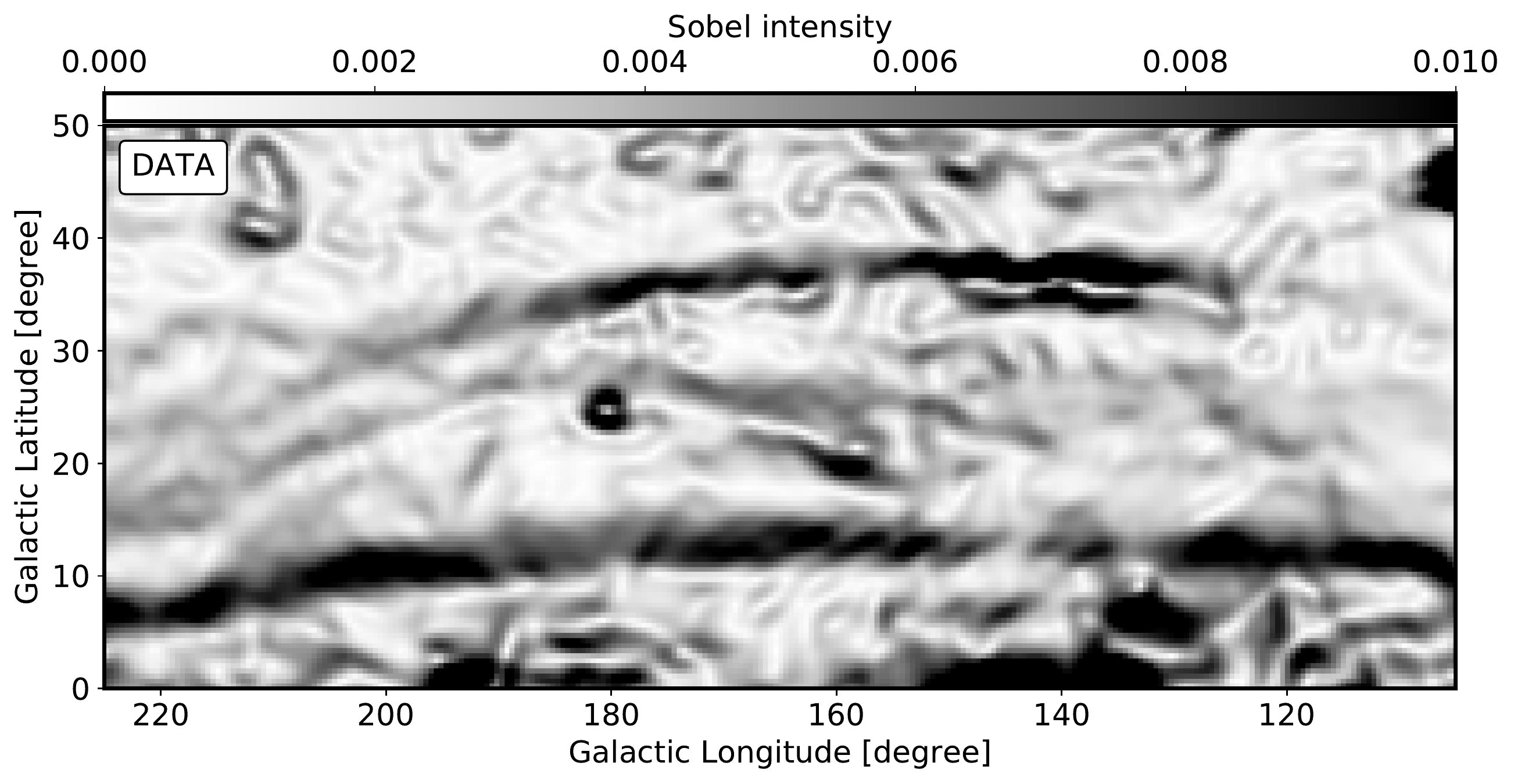}
    \includegraphics[width=0.7\textwidth]{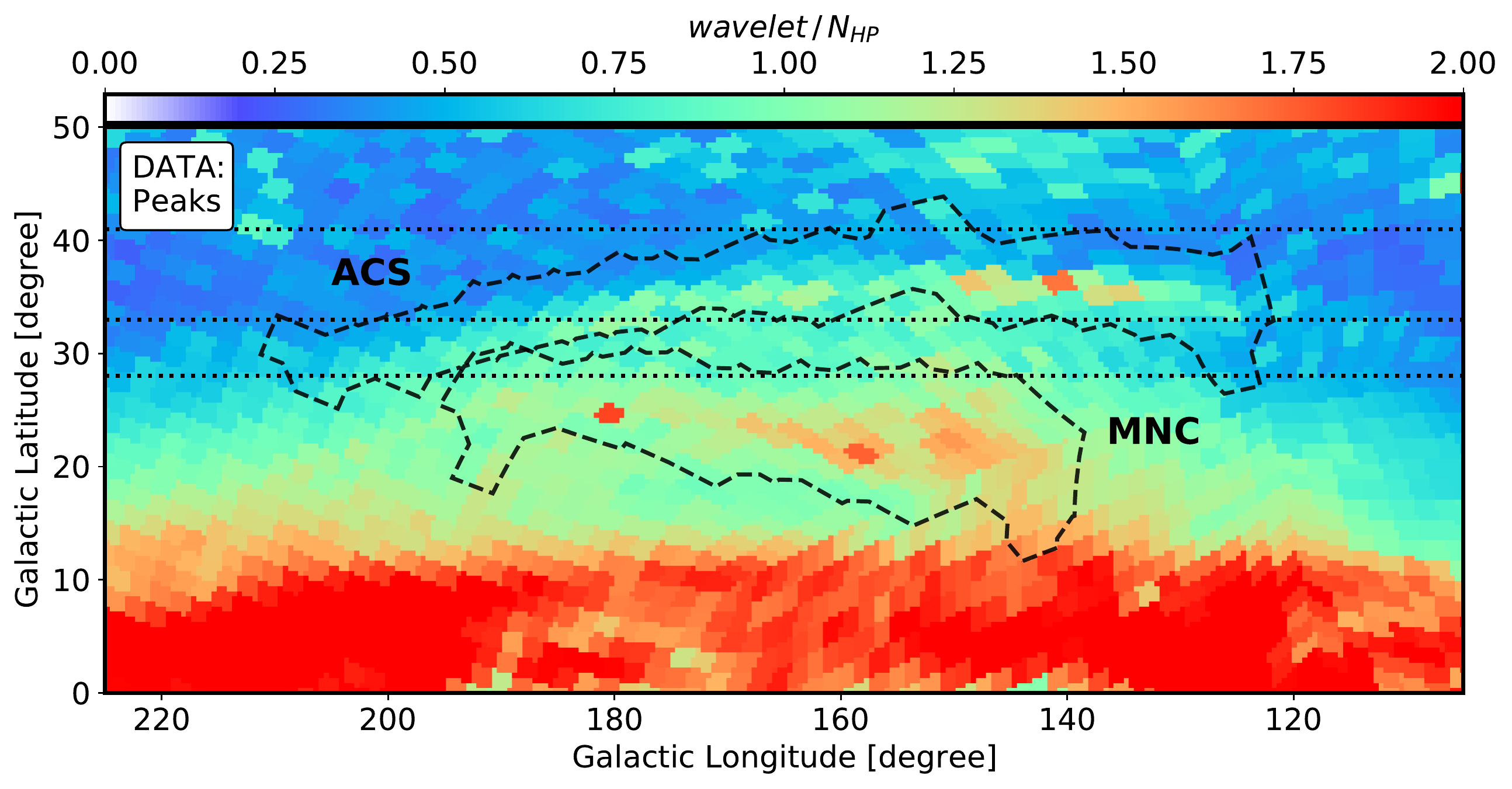}
    \caption{Zoom towards the anticentre region and definition of the patches. Top: Result of applying a Gaussian smoothing plus Sobel filter to Fig.~\ref{fig:allsky-wt} in the region: 110$^\circ$< l <220$^\circ$, -50$^\circ$< b <50$^\circ$. We use this map to isolate ACS and MNC. Bottom: $wavelet/N_{HP}$ (relative peak intensity) of the same region. The black contours delineate the regions that we have isolated according to the upper panel, while the dotted horizontal lines (b\,=\,[28$^\circ$,33$^\circ$,41$^\circ$]) give an approximate limit for the structures in latitude at l$\sim$180$^\circ$.}
    \label{fig:acs-sobel} 
\end{figure*}

\subsection{Monoceros and ACS}\label{sec:mnc&acs}

In Fig.~\ref{fig:acs-sobel} we present a zoom-in of Fig.~\ref{fig:allsky-wt} towards the anticentre and show our selection of the two structures that appear after colouring the sky according to the relative intensity of the dominant structure in proper motion. To build this selection in an objective manner, we first apply a Gaussian softening (two sigmas) of the 2D image to erase the \HP limits, and then apply a bi-directional \emph{Sobel} filter\footnote{Included in the \textit{Python} package {\normalfont\ttfamily Scikit-image} \citep{Scikit-image}} to reveal edges. By doing so, the two arches are cleanly separated at all longitudes. The final step is to select only the \HP whose \emph{Sobel} intensity is above a certain threshold (0.0035 for the bottom arch, 0.0040 for the top one). However, if we applied this selection blindly we would obtain a long list of \HP that comprises several structures. Instead, we first draw a rectangle around each arch in an appropriate coordinate system. This coordinate system, different for the bottom and top arch, is obtained by rotating the Celestial sphere with respect to the Galactic reference frame until the structure lies roughly flat at zero latitude in the new reference frame. The resulting final selections are the contours shown in the bottom plot of Fig.~\ref{fig:acs-sobel}. The structures can be seen to continue beyond the contours defined, especially for the feature at lower latitudes, but we focus on the regions where they are the most intense. We add three horizontal lines that represent an approximated latitude limit of each structure at $ell\sim$180$^\circ$.

By comparing the shape and location in the sky of these structures we note that they match with the MNC ring (bottom) and ACS (top) \citep[e.g.,][]{Newberg2002,Grillmair2006,Slater2014,Morganson2016}. The patches we obtain are also in good agreement with the regions delineated by \citet{Laporte2020} but are much more concise. In contrast to previous works, though, since we are not relying on counts but instead we detect these structures in relative intensity (Eq.~\ref{RI}), their morphology appears sharper and well defined. For instance, we observe a MNC structure that has a clear arch like shape\footnote{The strong red \HP at $(l,\,b)\sim$(180$^\circ$,\,25$^\circ$) is the globular cluster NGC 2419 which is far beyond MNC at a distance of $\sim$83\,kpc \citep{Forbes2008}.} extending from $\sim$120$^\circ$ to $\sim$230$^\circ$ in longitude, where it meets the disc at a latitude of $\sim$10$^\circ$. Nevertheless, we stress again that in the case of MNC, although the structure is physical, the sobel filter is enhancing the edge caused by the scanning law of \Gaiaf. In contrast, the ACS is thinner, stays above MNC for the ranges of longitudes where we detect it and has its strongest signal at $l\sim$140$^\circ$, where MNC has already almost merged with the disc. This is the most precise picture of the anticentre available to date, thanks to the introduction of kinematic information in the detection of these structures.

\begin{figure}
    \centering
    \includegraphics[width=\linewidth]{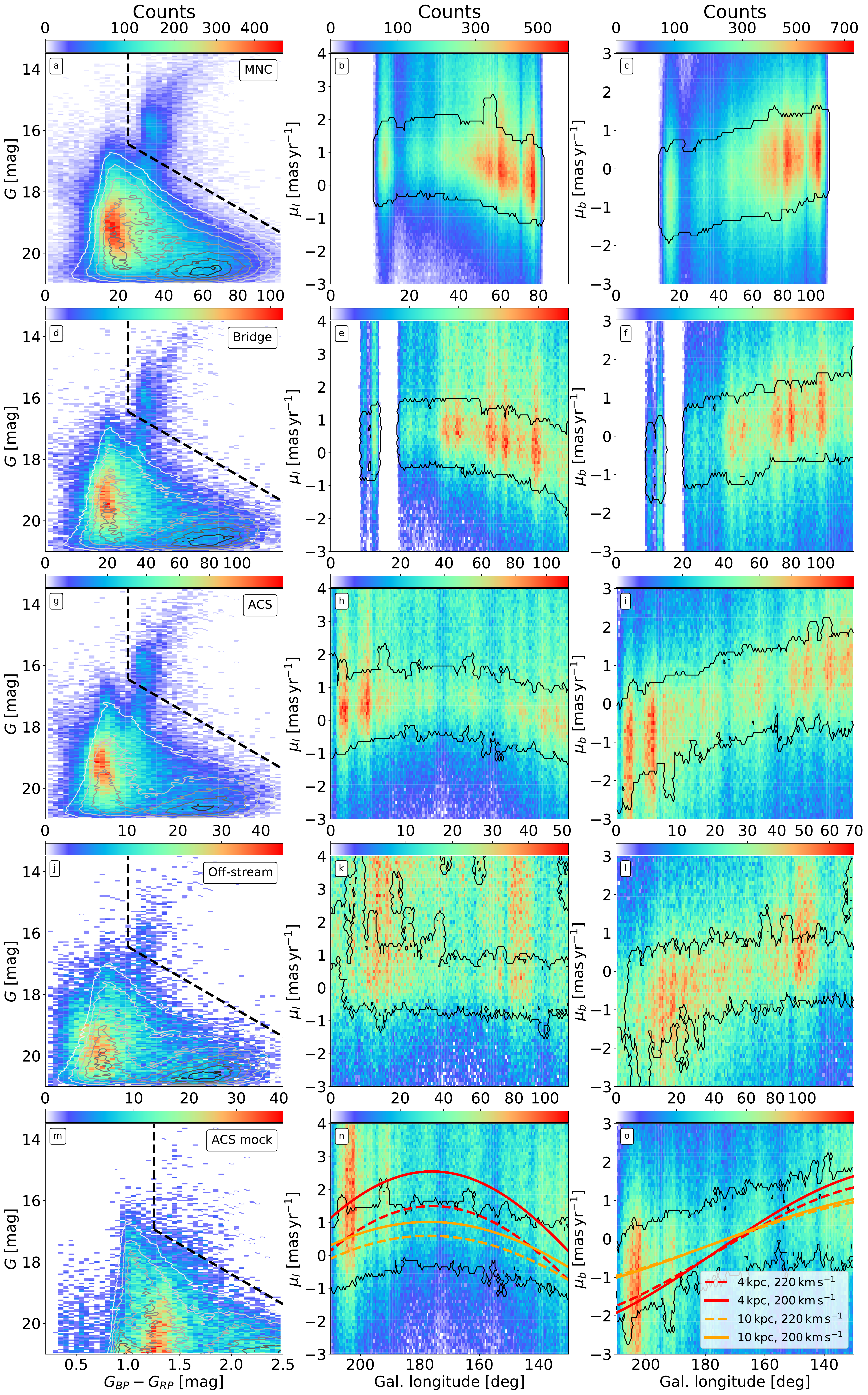}
    \caption{CMD (left) and proper motion trends with Galactic longitude in $\mu_l$ (middle) and $\mu_b$ (right) of each structure. The CMDs contain the histogram of the stars within the highest intensity proper motion peaks (peak stars) with grey contours on top that represent the CMDs of all the stars in the region. The black dashed lines represent our selection of Giant stars. The proper motion maps contain all the stars in the region, and the black line is the zero-contour of the peak stars, that is, the stars that fall within the highest intensity proper motion peak of their \HPf. First row: MNC. Second row: Bridge between MNC and ACS. Third row: ACS. Forth row: above ACS. Fifth row: Same region as ACS but for the mock particles, selecting the stars for the CMD according to the contours of panels \emph{h} and \emph{i}. In the bottom panel we also include the proper motions expected from a structure 30$^\circ$ above the plane at a given distance, 4 (red) or 10 (orange) kpc, rotating at a given velocity, 200 (solid) or 220 (dashed) km$\,$s$^{-1}$, but with no radial or vertical velocity.}
    \label{fig:patches_comparison}
\end{figure}

Once we have identified the MNC and ACS regions, we explore their kinematics and CMDs in more detail in Fig.~\ref{fig:patches_comparison}. Each row contains the results for different regions: the first row is MNC, the second corresponds to the list of \HP that fall between MNC and ACS (hereafter, the bridge), then the third is ACS and, finally, the fourth is a region above ACS. The last row is an example of what we would expect from a Galaxy with no substructure, obtained from a mock catalogue (Appendix \ref{app:mock}). With this exercise we can evaluate the continuity of these structures, compare their characteristics with nearby regions in the sky where we do not see an enhancement in relative intensity (c.f. Fig.~\ref{fig:acs-sobel}) and contrast them with the predictions of a MW model. In the first column of this plot we have aggregated all the stars that, in their respective \HPf, fall within the highest intensity proper motion peak. We refer to such stars as \emph{peak stars}. To show what we would see if we had not done this kinematic selection, the grey contours on top represent the CMD of all the stars of the region. The second and third columns contain, respectively, the trends of $\mu_l$ and $\mu_b$ with Galactic longitude for all the stars within the region. Here the black line encircles the stars selected kinematically, that is, the outer-contour of the volume that the peak stars occupy in this space. To provide some contrast with a fiduciary galaxy, in the bottom row we repeat the same process for the the particles in the mock catalogue that fall within the ACS footprint. In this case, the peak stars are selected according to the position and size of the peaks detected in the data, which is why the contours of panels h (i) and n (o) are so similar.

The first thing that we note is the presence of a Giant branch all the way from MNC to ACS, that disappears once we explore latitudes larger than $\sim$40$^\circ$. The fact that we see a well defined RC means that these stars share a similar distance which, based on  their magnitudes ($\sim$15.5\,mag) and Galactic latitudes (b\,>\,15$^\circ$), puts them well above the mid-plane of the Galaxy (z\,>\,2\,kpc) at a height larger than the scale height of even the thick disc. If we compare  the observed CMDs with the CMD of the mock catalogue, we note that we do not expect many stars in this region of the diagram as the nearby giants have been already removed with the cut in parallax (these are bright enough to have a reliable parallax) and the farther ones are not in the model since there are not many stars at such heights/distances.

We also note, accompanying the Giant branch, a dense clump of stars appearing in panels a-d-g which is bluer than the MS of the disc seen in the mock. \citet{Newberg2002} already reported that the main sequence turn off of MNC was bluer than the thick disc and we detect the same behaviour for the stars in the peaks found within MNC and ACS. This group of stars is consistent with being the MS of an isochrone containing the RC discussed above. The rest of the stars that fall outside said isocrhone seem to follow the contours of the CMD obtained with all the stars (no kinematic selection,  grey contours in those panels), and are most likely nearby, dwarf field stars that overlap with these structures in the proper motion plane. In contrast to previous works \citep[e.g.][]{Newberg2002,Ivezic2008,Xu2015,Thomas2019}, where they detect an overdensity in counts for a given population, MS, MSTO, blue stragglers or M-giants, here we have unveiled the whole sequence in the CMD by performing a blind kinematic selection of the stars instead.

In the bottom row of Fig.~\ref{fig:patches_comparison} we have included a few curves that represent the proper motion that \Gaia would measure for a star at given distance if it only had azimuthal velocity\footnote{Here, we simply used the analytical expressions that transform the velocity of a star at a given position ($l$, $b$ and distance) and that has only rotational velocity to proper motions in $\mu_l$ and $\mu_b$. For the position and velocity of the Sun with respect to the GC we have used $R_\odot$ = 8.178\,kpc \citep{Gravity2019} and $V_\odot$ = [11.1,\,248.5,\,7.25]\,km$\,$s$^{-1}$ \citep{Schonrich2010,Reid2020}. In all the cases we keep the latitude constant to 30$^\circ$ since the lines are used to compare with the tracks obtained in the ACS region, but we obtain similar results for the MNC region.}. In orange (red), this distance is 10\,kpc (4\,kpc) and in solid (dashed) the rotation velocity is 200$\kms$ (220$\kms$). A structure that is too near, like the brown curve, does not match the contour delineated by the peak stars (black contours), while a structure that does not rotate, as could be the halo, would have to be at a distance larger than 50\,kpc in average to fall within the black lines. And even then, its shape would not be compatible with the data. While we note that other combinations of distances and velocities could produce a similar shape (even if the result is not physically supported), the dashed red line shows a good agreement with our observations and corresponds to a structure at $\sim$10\,kpc rotating slightly slower than the disc. In other words, the peak stars in ACS (the same applies to MNC) have proper motions that change with Galactic longitude in a way that is compatible with a structure at a distance of $\sim$10\,kpc rotating at a speed similar to the disc or slower, in agreement with the analysis by \citet{deBoer2018}.

By comparing the CMDs inside and outside the patches defined in Fig.~\ref{fig:acs-sobel}, it is clear that, even though the MS of these structures is the dominant fraction, by focusing only on the giants we can gain contrast with the MW foreground. Therefore, we introduce another tag, apart from the one that we have already been using to separate stars inside and outside the proper motion peaks. The stars will be called giants whenever they are redder than $G_{BP}-G_{RP}\,>\,$1\,mag and their apparent magnitude smaller (brighter) than the line:
\begin{equation}\label{eq:giants}
    G < 1.95(G_{BP}-G_{RP})+14.50,
\end{equation}
\noindent where we have used the slope calculated in \citet{RomeroGomez2019} to follow the extinction vector, and the zero-point is adjusted by eye to reduce the contamination from the disc while preserving the RC as much as possible. We note, however, that we are still selecting some faint, red dwarfs at all latitudes, the great majority of which are \emph{not} classified as peak stars (for sources redder than 2\,mag in $G_{BP}-G_{RP}$ and $G$>14\,mag, only $\sim$100 out of $\sim$8000 in MNC and $\sim$30 out of $\sim$4000 in ACS). This means that the giants tagged also as peak stars are more likely to be true giants, whereas field stars tagged as giants have a larger probability of being nearby red dwarfs.

\begin{table*}
    \begin{center}
    \caption{MNC stars classified both as peak and giants (top 2 rows). The first column contains the \begin{tt}source\_id\end{tt} followed, in columns two and three, by the right ascension and declination of the star. Columns four to seven contain the proper motions (ICRS) and the corresponding uncertainties. The eight and ninth are the apparent magnitude in the $G$ band and the \Gaia colour $G_{BP}-G_{RP}$, respectively. Then, the Galactic coordinates, $\ell$ and $b$ are given in columns twelve and thirteen. Finally, in the last column, we give the absorption in the $G$ band (see text).}\label{tab:sample_mnc}
    \begin{tabular}{cccccccccc}
    \hline\hline
    source\_id &  ra & dec & $\mu_{\alpha*}$&  $\sigma_{\mu_{\alpha*}}$ &     $\mu_{\delta*}$ &  $\sigma_{\mu_{\delta}}$ & $G$ & $G_{BP}-G_{RP}$ & A$_G$\\
    & [$^\circ$] & [$^\circ$] &
    [mas\,yr$^{-1}$] &  [mas\,yr$^{-1}$]  & [mas\,yr$^{-1}$]  &  [mas\,yr$^{-1}$] & [mag] & [mag] & [mag]\\
    \hline
      926626863262740096 & 116.29 & 43.76 & 0.0070 & 0.1237 & -0.9832 & 0.0973 & 16.40 & 1.20 & 0.10 \\
926656550076613376 & 116.15 & 44.00 & -0.6643 & 0.0817 & -0.6905 & 0.0564 & 15.66 & 1.11 & 0.10 \\
    \hline
    \hline
    \end{tabular}
    \end{center}
\end{table*}

\begin{table*}
    \begin{center}
    \caption{Same as Table~\ref{tab:sample_mnc} but for ACS (top 2 rows).}\label{tab:sample_acs}
        \begin{tabular}{cccccccccc}
    \hline\hline
    source\_id &  ra & dec & $\mu_{\alpha*}$&  $\sigma_{\mu_{\alpha*}}$ &     $\mu_{\delta*}$ &  $\sigma_{\mu_{\delta}}$ & $G$ & $G_{BP}-G_{RP}$ & A$_G$\\
    & [$^\circ$] & [$^\circ$] &
    [mas\,yr$^{-1}$] &  [mas\,yr$^{-1}$]  & [mas\,yr$^{-1}$]  &  [mas\,yr$^{-1}$] & [mag] & [mag] & [mag]\\
    \hline
      703742318576330496 & 128.16 & 26.59 & -0.2006 & 0.0740 & -0.7579 & 0.0510 & 15.12& 1.23 & 0.09 \\
      703750633632989056 & 128.28 & 26.72 & -0.1967 & 0.1102 & -0.3101 & 0.0762 & 16.26 & 1.18 & 0.11 \\
    \hline
    \hline
    \end{tabular}
    \end{center}
\end{table*}

\begin{figure}
    \centering
    \includegraphics[width=\linewidth]{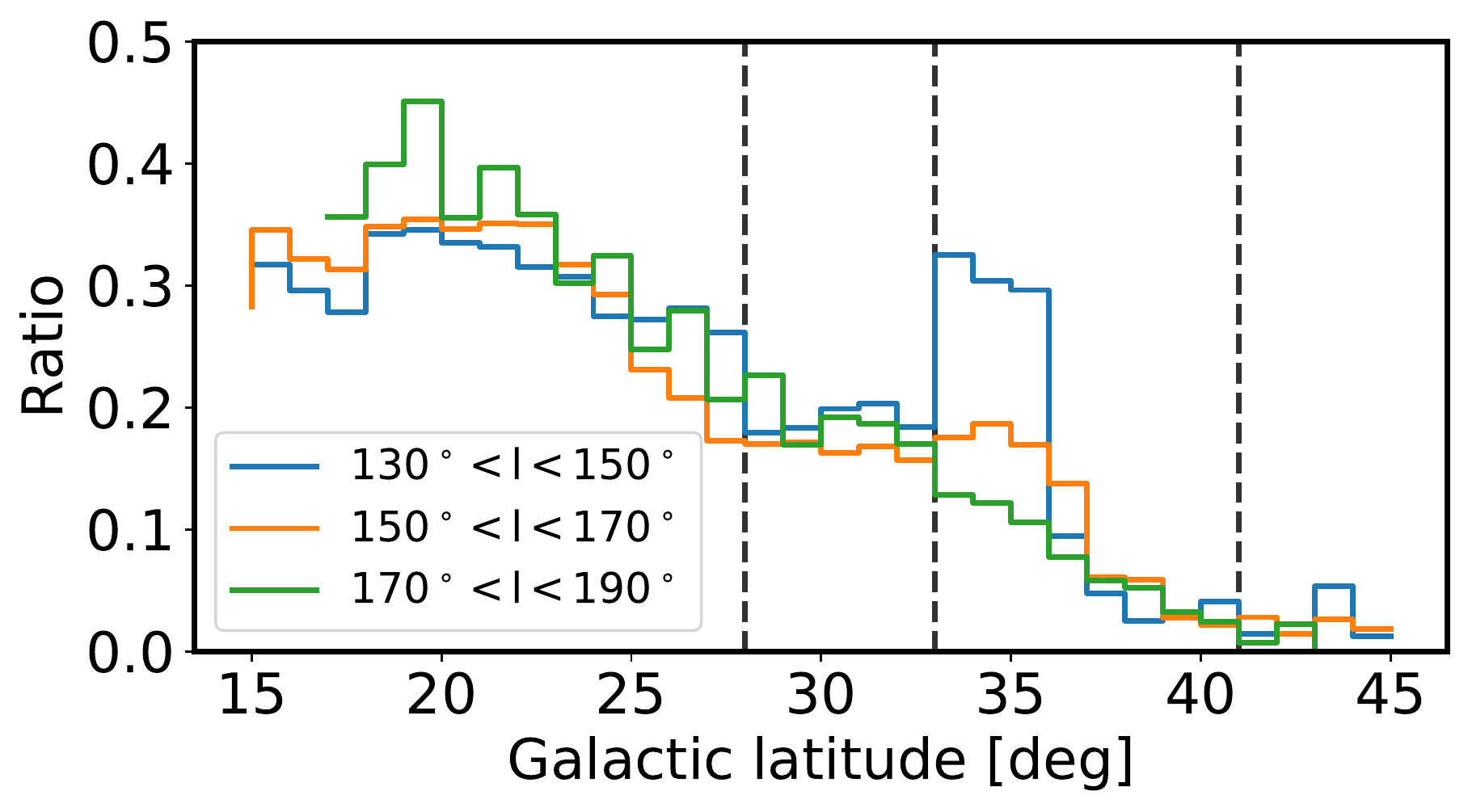}
    \caption{Ratio of giants in the peak compared to all the giants as a function of Galactic latitude for three different ranges in longitude: 130$^\circ$ < l < 150$^\circ$ (blue), 150$^\circ$ < l < 170$^\circ$ (orange), and 170$^\circ$ < l < 190$^\circ$ (green). The vertical lines give an orientation of the end of each structure with Galactic latitude (c.f. Fig~\ref{fig:acs-sobel}). A sudden increase of the ratio can be seen in the part where ACS is the more intense.}
    \label{fig:bVSratio}
\end{figure}

Figure~\ref{fig:bVSratio} shows the fraction of giants inside the peaks (i.e., stars tagged as giants and peak stars simultaneously) with respect to all the stars tagged as giants as a function of latitude. Given that our classification is rather rough, we should treat these as simple estimates and focus on the trends instead. What we observe is that the parts where MNC and ACS have the strongest signal in relative intensity (c.f. Fig.~\ref{fig:acs-sobel}) coincide with the regions where this ratio is the highest. We have already seen that the relative intensity of ACS decreases with with longitude, and here we note that the ratio of giants also diminishes moving from one curve to the other. We also observe that the bridge keeps a constant ratio, showing that it is just the region where the tail of the two structures overlap. Finally, we note that our patch around ACS is too broad, as the ratio drops abruptly at $b\sim$37$^\circ$, coinciding with the place where we observe a discontinuity in the kinematics\footnote{We have checked that, indeed, the trends in proper motion as a function of latitude also suffer a sudden change at around $\sim$37$^\circ$, as can be seen in Fig.~\ref{fig:allsky-pm}.}. 

Tables~\ref{tab:sample_mnc} and \ref{tab:sample_acs}, available online, contain the list of sources classified simultaneously as peak stars and as giants for, respectively, MNC (10\,079 sources) and ACS (2\,104 sources).

\subsection{Anticentre region: North vs South}\label{sec:northVSsouth}

\begin{figure}
    \centering
    \includegraphics[width=\linewidth]{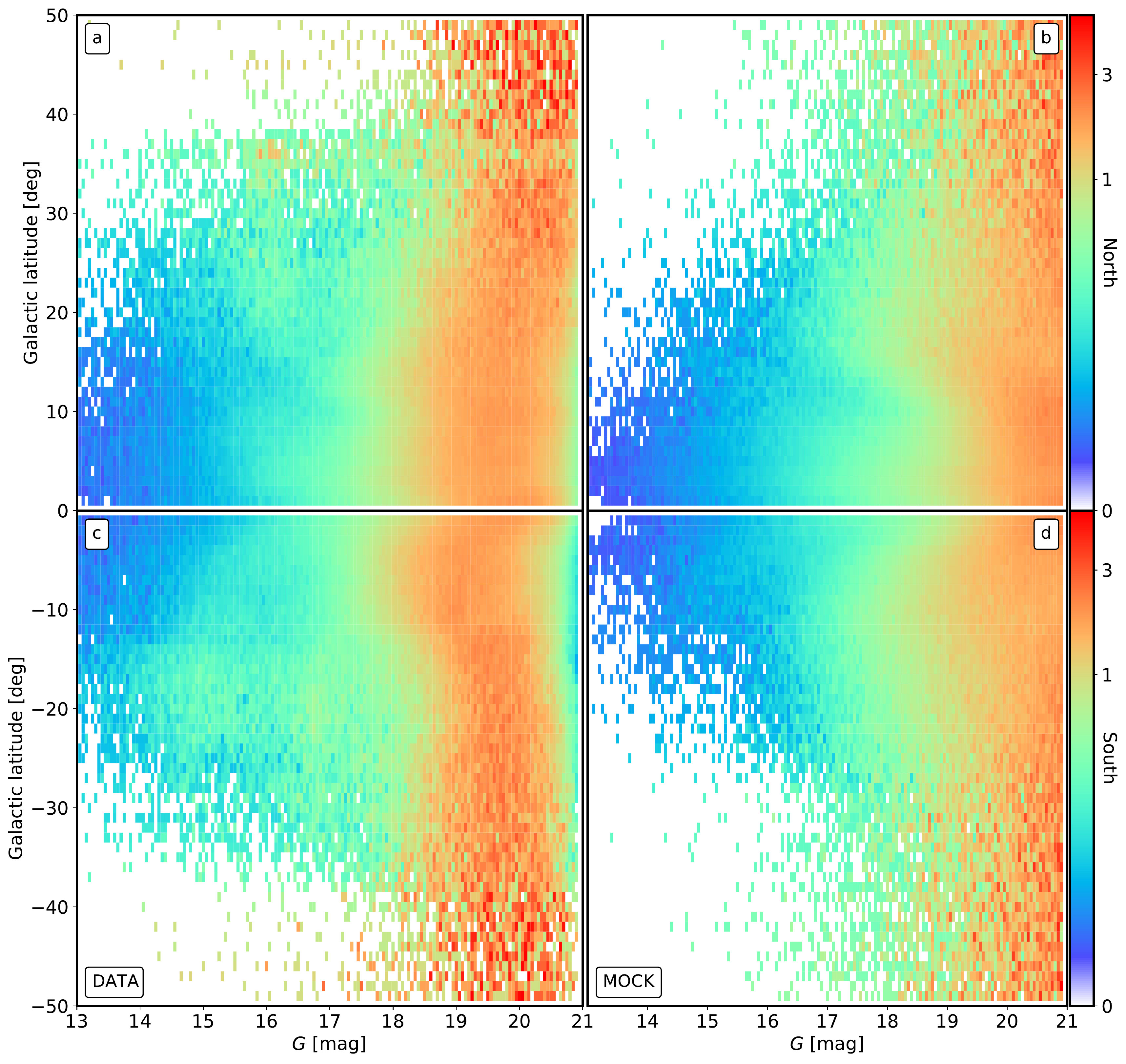}
    \caption{Apparent magnitude as a function of Galactic latitude for the stars in the peaks with 130$^\circ$ < l < 150$^\circ$. The histograms are normalised such the sum of all pixels in a given bin of $b$ adds up to unity. A conspicuous overdensity of stars can be clearly seen appearing at $G\sim$16\,mag and extending from 15$^{\circ}$\,<$b$\,<40$^{\circ}$. Left: Data. Right: Mock.}
    \label{fig:bVSgmag_130-150}
\end{figure}

Above, we have proven that our methodology can detect kinematic substructures and isolate them effectively from the rest of the disc. Also, that the giants inside the proper motion peaks are good tracers of MNC and ACS since the contamination in that region of the CMD is expected to be very low once we have filtered by parallax and kinematics. Therefore, we can use the location of the RC to trace the structures in physical space. To do so, we now explore the changes in apparent magnitude of the stars that we have tagged as peak stars, for different ranges of longitude, both in the north and south hemispheres. 

Figure~\ref{fig:bVSgmag_130-150} shows the distribution of apparent magnitudes of peak stars with respect to Galactic latitude using a histogram normalised by bins of $b$, for the range 130$^\circ$ < l < 150$^\circ$. This figure confronts the data (top) with the expectations from the mock (bottom). In this case, the particles selected in the mock correspond to the peaks that are detected in the mock itself (in contrast to panel m of Fig.~\ref{fig:patches_comparison} where we used the peaks detected in the data). We find an overdensity of stars in the north at a magnitude $\sim$16 that corresponds to the RC seen in Fig.~\ref{fig:patches_comparison}. It is most intense above $b>$30$^\circ$ and corresponds to the ACS. We see it extending rather continuously down to $b\sim$10$^\circ$ where it merges with the disc, following an arch that is compatible with the increase in extinction.

\begin{figure*}
    \centering
    \includegraphics[width=\linewidth]{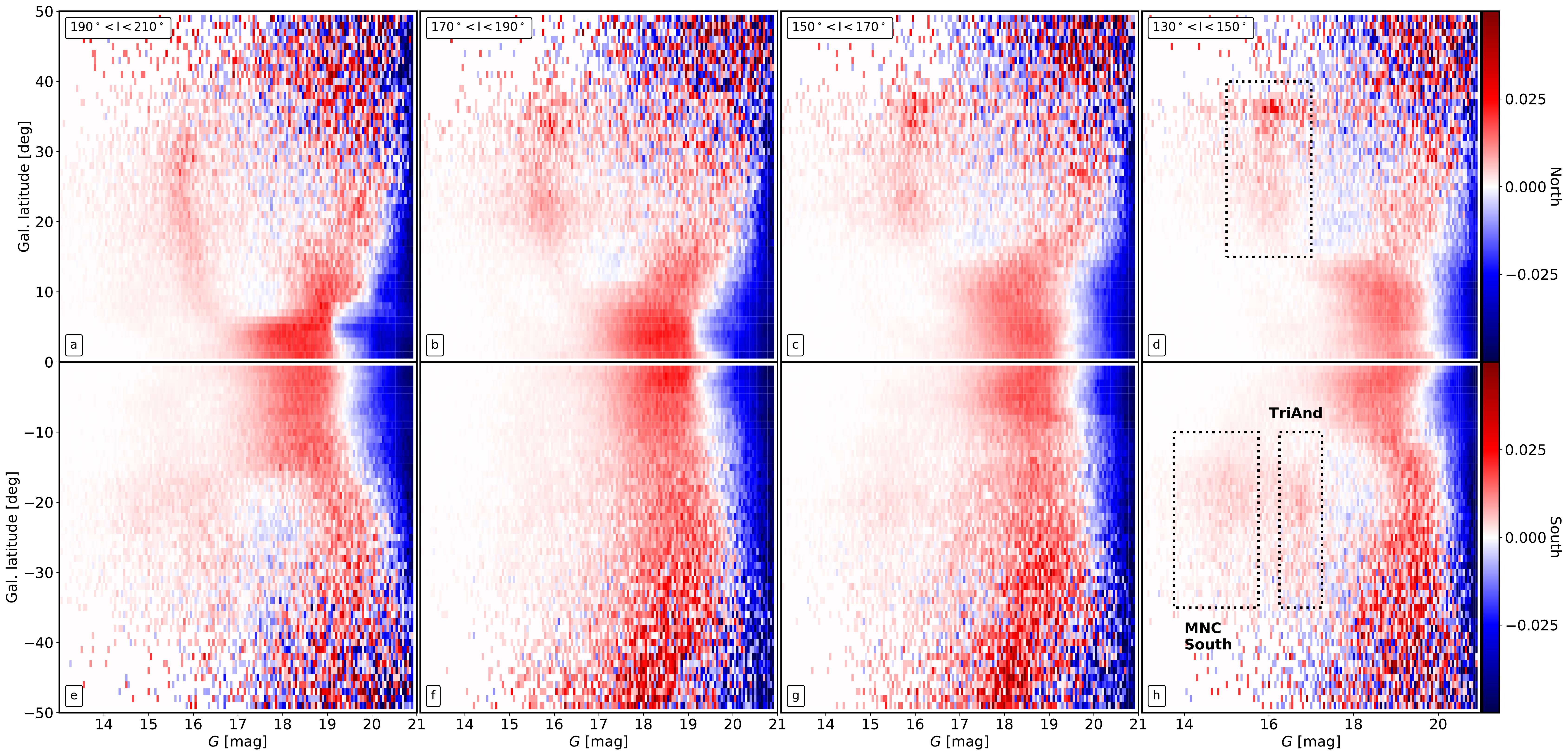}
    \caption{Differences between the data and the mock in the plane of apparent magnitude against Galactic latitude of counts in the peak stars, for different bins in longitude. The plots are obtained by subtracting the counts from the top and bottoms panels of Fig.~\ref{fig:bVSgmag_130-150}, and similar panels for the other ranges of longitude, which had been previously normalised by bins of latitude. From right to left: 130$^\circ$ < l < 150$^\circ$, 150$^\circ$ < l < 170$^\circ$, 170$^\circ$ < l < 190$^\circ$, and 190$^\circ$ < l < 210$^\circ$.}
    \label{fig:bVSgmag_diff}
\end{figure*}

In the south, we observe an excess of bright stars ($G<$17$\,$mag) at latitudes between 15$^\circ$ and 25$^\circ$ with respect to the mock. Interestingly enough, the intensity maps (Fig.~\ref{fig:allsky-wt}) do not show an enhancement as is the case for the north, not even compared to the mock map (Fig.~\ref{fig:allsky-wt_mock}). Based on their apparent magnitudes and location in the sky, it is very likely that these stars form the diffuse stellar population detected by \citet{Ibata2003} and that is sometimes called MNC South. This is also the region where the TriAnd overdensities have been reported \citep{Majewski2004, Rocha-Pinto2004, Martin2007} and, as we show below, we do detect it with our method. A detailed study of these structures and a comparison with previous studies \citep[e.g., Fig.\,4 from][]{Perottoni2018} is out of the scope of this work, but we will revisit it once the eDR3 \Gaia data release \citep{Brown2019} is made public and we have more and better astrometric data.

In Fig.~\ref{fig:bVSgmag_diff} we now present the difference between the data and the mock in the same plane of apparent magnitude against Galactic latitude, for different bins in longitude. From right to left, these are: 130$^\circ$ < l < 150$^\circ$, 150$^\circ$ < l < 170$^\circ$, 170$^\circ$ < l < 190$^\circ$, and 190$^\circ$ < l < 210$^\circ$. In panels (d) and (h) we see the subtraction of the left panels of Fig.~\ref{fig:bVSgmag_130-150} from the right panels (i.e., data minus mock after properly normalising the histograms). As already mentioned, we see two distinct overdensities in the south, a bright (14$<G<$15.5\,mag) one corresponding to MNC south and a fainter one (G$\sim$17\,mag) corresponding to TriAnd, at a magnitude consistent with the most recent determinations of its heliocentric distance \citep[e.g.,][]{Bergemann2018}. These features cannot be seen so clearly at other longitudes except for panel (e) where the diffuse overdensity appears again (latitudes between $\sim$20$^\circ$ and $\sim$30$^\circ$, brighter than $\sim$17\,mag). While we cannot discard that the southern structures are indeed discontinuous, taking into account the corrugations in the disc reported by \citet{Xu2015} the most likely scenario is that we can only detect them with our method where the extinction is low enough. With the next \Gaia releases we will be able to assess better their continuity. 

In the north (top panels), we note that ACS decreases its intensity and shifts to lower latitudes when we move towards the third quadrant of the Galaxy, as we see also in Fig.~\ref{fig:acs-sobel}. In the intermediate panels (b and c) we observe two concentrations at different latitudes, but similar apparent magnitudes, corresponding to MNC and ACS, whose tails overlap forming the bridge that we mentioned above. More importantly, we see MNC extending more and more towards lower latitudes, keeping roughly the same apparent magnitude throughout. This is interesting as MNC is usually hard to trace so deep into the disc due to the foreground stars. And yet, in panel (a), using our kinematic selection its RC can be traced down to a latitude $\sim$5$^\circ$.  

\begin{figure}
    \centering
    \includegraphics[width=\linewidth]{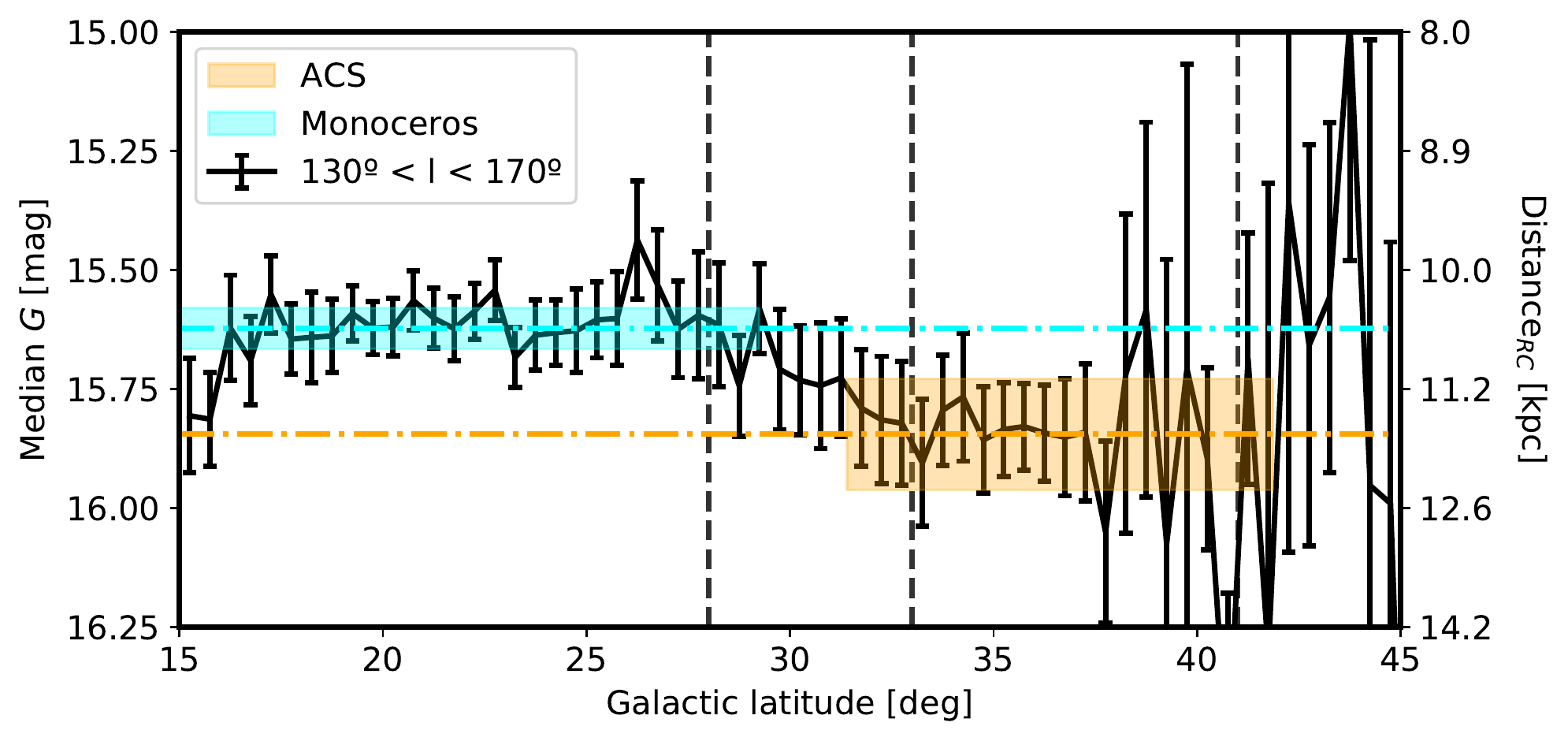}
    \caption{Apparent magnitude of giants peak stars in the anticentre region (130$^\circ$ < l < 170$^\circ$) as a function of Galactic latitude, after correcting for extinction (see text). The error bars denote the 1$\sigma$ uncertainty on the median computed as $\sigma\sqrt{\frac{\pi}{2N}}$, where $\sigma$ is the standard deviation of the apparent magnitude in the bin. Vertical lines represent the approximate limits of each structure in that range of Galactic longitudes (see Fig.~\ref{fig:acs-sobel}), and the right axis represents the distance to a RC star with the apparent magnitude shown in the left axis. The horizontal lines correspond to the median $G$ magnitude for the giants peak stars within MNC (cyan) and ACS (orange). The shaded areas contain the $\pm$3$\sigma$ interval of uncertainty on the median and they extend from the minimum to the maximum latitude of the peak stars within each patch (the vertical dashed lines serve only as an orientation). As can be seen, the ACS is fainter than MNC and this translates to a difference in distance of $\sim$1\,kpc. }
    \label{fig:bVSgmag}
\end{figure}

By measuring the median $G$ for the Giant stars only, selected according to Eq.~\ref{eq:giants}, we can investigate the relative distance of these structures. To do so, however, since the latitudes probed by MNC and ACS change with longitude, we focus only in the range 130$^\circ$ < l < 170$^\circ$ where they remain rather flat and the bridge is quite wide (see Fig.~\ref{fig:acs-sobel}). The median of the $G$ magnitude and the associated one sigma interval of uncertainity at different bins in latitude is shown in Fig.~\ref{fig:bVSgmag} for the Giant stars inside the proper motion peaks. To compensate for the effects of extinction, we have first corrected the apparent magnitude using the $G_{BP}-G_{RP}$ colour of each source individually and the prescription detailed in Appendix~A of \citet{Ramos2020}. What we observe is that, below $b\sim$28$^\circ$ where we identify MNC, the RC is brighter than above $b\sim$31$^\circ$ where ACS begins. Since we have used the integrated extinction up to infinity \citep{Schlegel1998}, the separation that we observe is an upper limit: if we assume that the extinction applied to ACS is correct, since it is at a higher latitude, then MNC could actually be less extincted than assumed, and therefore be intrinsically fainter than the value we are recovering. Nevertheless, since both structures are quite far, this effect should be small and we can safely conclude that ACS is farther away than MNC. 

To be more quantitative, we measure now the median $G$, corrected for extinction, for all the peak giant stars in this longitude range.  
The result is the shaded areas shown in Fig.~\ref{fig:bVSgmag} where we can clearly see that ACS is, once we convert the difference in magnitude to distance, roughly 1\,kpc farther away than MNC, with a discrepancy of more than 3$\sigma$. Also, we have estimated the median distance to each of the two structures. In doing so, we assume that the median apparent magnitude measured (horizontal lines in Fig.~\ref{fig:bVSgmag}) corresponds to the magnitude of the RC. By imposing that the absolute magnitude of a RC stars is $M_G$=0.495\,mag \citep{RuizDern2018}, we obtain the following median distances and their statistical uncertainties\footnote{Here, the dominant source of uncertainty is the systematic errors, which are not included in the error bars given. The more important ones are i) the assumption that the median magnitude is the magnitude of the RC, and ii) not using 3D extinction maps but instead correcting with the integrated extinction to infinity. Other sources of systematic uncertainty are the error on the absolute magnitude of the RC, contamination from stars that are not giants, or errors in the extinction map.}: D$_{MNC}$\,$\sim$\,10.6$\pm0.1$\,kpc, and D$_{ACS}$\,$\sim$\,11.7$\pm0.2$\,kpc.

Nevertheless, without a precise calibration of each individual star, and its extinction, we cannot investigate the changes in distance with longitude and latitude which is key to reveal the 3D shape of these structures. We make a first attempt to study the distribution of the structures we detect along the line of sight by cross-matching the peak stars with StarHorse \citep{Anders2019}, a catalogue of Bayesian derived astrophysical parameters obtained from the photometry of \Gaia, Pan-STARRS1, 2MASS, and AllWISE combined. We download all the stars in the anticentre (100$^\circ$ < l < 260$^\circ$ and -60$^\circ$ < b < 60$^\circ$) with {\normalfont\ttfamily{SH\_OUTFLAG}} equal to "00000", as recommended in \citep{Anders2019}, and with a distance (50th percentile) less than 20\,kpc. From the 13\,098\,038 peak stars in the north, we find 429\,565 in StarHorse. In the south, the cross-match returns 514\,167 stars out of the 13\,669\,647 peak stars. Most of them, however, are faint dwarfs found at low latitudes, closer than 10\,kpc, whose parallax quality is not good enough to discard them with the filter presented in Sect.~\ref{sec:data&methods}. If instead we restrict ourselves to the giant peak stars, then we find 378\,955 (out of 1\,449\,250) in the north and 458\,403 (out of 1\,286\,132) in the south. 

\begin{figure*}
    \centering
    \includegraphics[width=\linewidth]{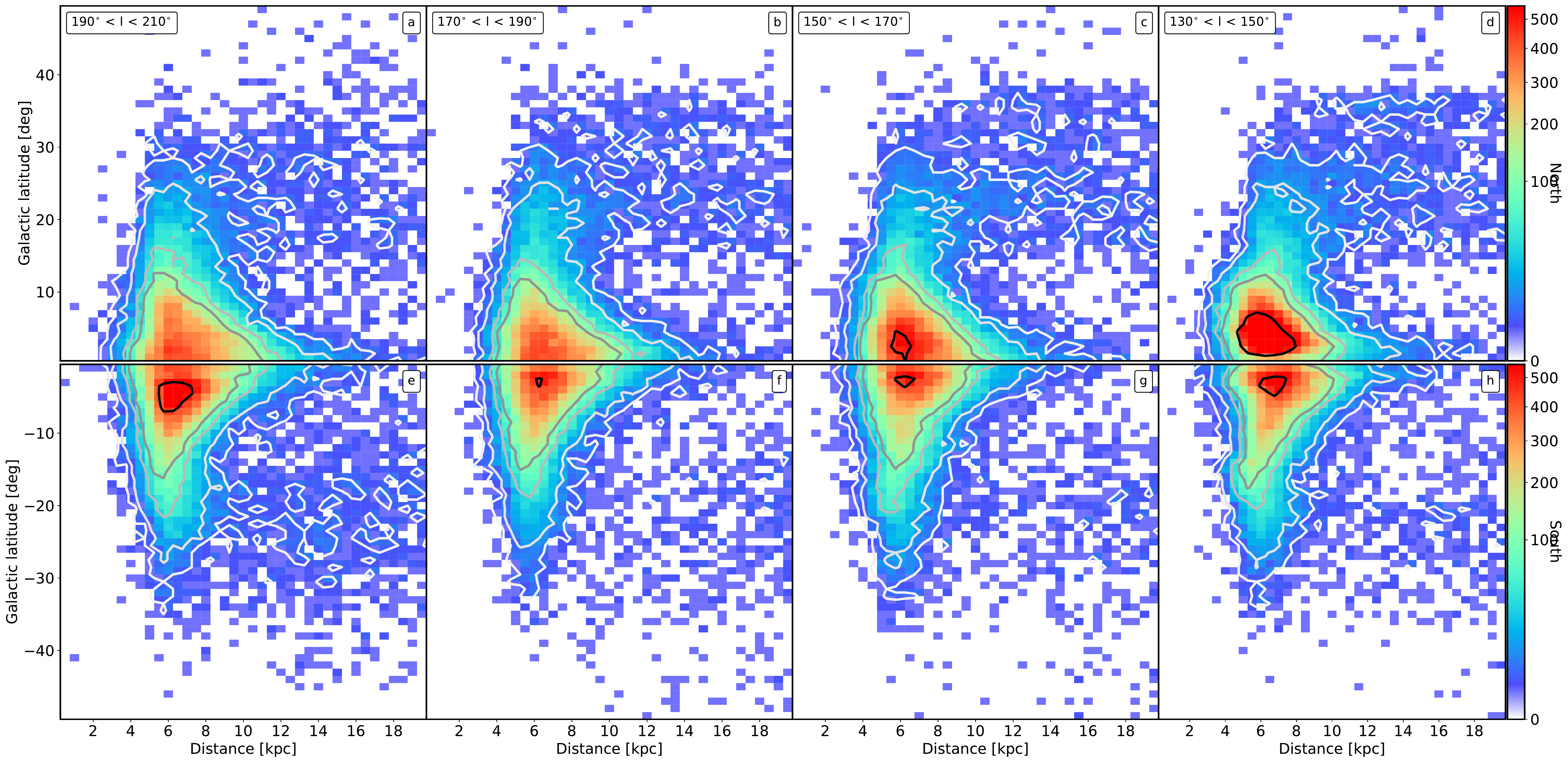}
    \caption{Distance from StarHorse as a function of Galactic latitude for different slices in Galactic longitude, only for the giants stars within the peak. From right to left: 130$^\circ$ < l < 150$^\circ$, 150$^\circ$ < l < 170$^\circ$, 170$^\circ$ < l < 190$^\circ$, and 190$^\circ$ < l < 210$^\circ$.}
    \label{fig:bVSdist_data}
\end{figure*}

Figure~\ref{fig:bVSdist_data} shows the distribution of StarHorse distances as a function of Galactic latitude for the four ranges of longitude explored above. The first thing we note is the effect of our selection function as the nearby giants are missing and a wall of stars at a distance of $\sim$6\,kpc is formed. The tails extend up to $\sim$15\,kpc, point beyond which StarHorse distance uncertainties become too large. Compared with the corresponding figure for the mock (Fig.~\ref{fig:bVSdist_mock}), where we see the disc extending much farther away, we note a clear excess of stars in the data at latitudes larger than 20$^\circ$ and at a distance of >7\,kpc. We associate these to MNC and ACS. In the right column, in the range of longitudes where ACS is more intense, we see it clearly separated from MNC and slightly farther away. As we shift our view towards the third Galactic quadrant, these structures recede, becoming less prominent and shifting to lower latitudes (as we showed above). MNC covers a large range of latitudes and connects smoothly with the disc, but the lack of stars and the uncertainties prevent us from determining if there is a distance gradient with latitude or not. 

The south does not show the same structures as the north but we note that, at least for panels (f) and (g), the extinction is higher than in the north, which could block our line of sight. We note that we do not recover so clearly the structures detected in panel (h) of Fig.~\ref{fig:bVSgmag_130-150}, probably due to low statistics. However, we do observe an increase of stars in panel e in the form of a diffuse distribution of distant stars at latitudes between -20$^\circ$ and -30$^\circ$, coinciding with the location of the structure S200-24-19.8 reported in \citet{Newberg2002} and also the detection by \citet{Xu2015} that they associate with the aforementioned TriAnd.

\subsection{RR Lyrae to M giant ratio}\label{sec:f_rr2m}

With a kinematically selected sample of stars for both MNC and ACS, we can now check if their population is consistent with having been born in an extragalatic system or not. All known MW dSphs have a large fraction of RR Lyrae stars \citep{Vivas2006} and a low one of M giants \citep{Price-Whelan2015}, whereas the opposite happens with the Galactic discs. Hence, we will follow \cite{Price-Whelan2015}, where they used the ratio between the number of RR Lyrae and M giant stars in TriAnd \citep{Rocha-Pinto2004, Martin2007} to argue that these structures were probably disc stars kicked-out by an external perturbation. For that, we need to estimate the number of RR Lyrae and M giants within MNC and ACS. Based on previous studies, we expect a low number of RR Lyrae in these structures \citep{Kinman2004} but a large number of M-giants \citep[e.g.,][]{Rocha-Pinto2004}. Now the question is how low is the ratio between the two populations.

Starting with the RR Lyrae, we use the catalogue described in \cite{Mateu2020} which combines the VariClassifier and Specific Objects Studies (SOS) catalogues from Gaia DR2 \citep{Holl2018,Clementini2019} with the ASAS-SN-II catalogue \citet{Jayasinghe2019a},  providing optimal completeness at the bright end ($G<15$). 
We select only those sources that fall within the sky patches defined in Fig.~\ref{fig:acs-sobel}, for a total of ~900 (~800) RR Lyrae in MNC (ACS), of which 253 stars fall within the one standard deviation range around the median distance in the case of MNC (6.7 to 16.7\,kpc) and another 253 for ACS (7.8 to 17.7\,kpc). Of these, only 12 (6) are also consistent with the kinematic signature we have detected, in the space of l-b-pmra-pmdec. Based on these results, and after correcting for the completeness of the RR Lyrae catalogue \citep[estimated at ~80\% at these magnitudes in][]{Mateu2020}, we conclude that, at most, MNC has 15 
RR Lyrae and ACS, no more than 8.
In parallel, we cross-match the list of RR Lyrae with our sample to see the effect that the cut in parallax\footnote{The effect of the cut in colour is negligible} has and we observe that, from the 253 (253) stars that we had, only 123 (123) remain in MNC (ACS). Finally, if we now keep only those classified as peak stars (i.e., probable MNC/ACS members) we find only 1 and 2 RR Lyrae for, respectively, MNC and ACS. These figures are lower and, even if we correct by the 48.6\% reduction in completeness caused by the cut in parallax, the maximum amount of RR Lyrae in MNC (ACS) would be 3 (5).

On the other hand, the M giants are much more numerous. We use the official \Gaia cross-match with 2MASS and the selection proposed by \cite{Majewski2003} to obtain the corresponding $G_{BP}-G_{RP}$ colour cut necessary for MNC and ACS, finding that we can select confidently M giants in our sample of candidates with the following limits: $G<15.5$mag and $G_{BP}-G_{RP}>1.5$mag. These cuts result in a total of 959 M giants for MNC and 155 for ACS. Of course, these values are not corrected for the completeness of the sample, as opposed to those corresponding to the RR Lyrae. Therefore, if we take the maximum number of RR Lyrae that these structures can have, the ratio $f_{RR:MG}$ that we provide becomes a strict upper-limit: $f_{RR:MG} < 1.5\%$ for MNC, $f_{RR:MG} < 5.2\%$ for ACS.

The fractions obtained are consistent with previous independent estimates \citep{Sheffield2018} and are totally compatible with a stellar population of the MW alpha-poor disc and much lower than the expected values for extragalactic systems like Sagittarius or the LMC, for which we expect a fraction $\sim$50\% \citep{Price-Whelan2015}. In fact, it is hard to reconcile these values with the hypothesis that MNC and ACS are tidal tails of an accreted satellite. It would require either a young system that managed to reach high metallicities (perhaps formed from already metal-enriched gas), or the effect of a dynamical mechanism that could segregate RR Lyrae from giants within the stream, something that we do not observe in other streams like the Sagittarius stream \citep[e.g.,][]{Antoja2020,Ibata2020,Ramos2020}.

\section{Discussion}\label{sec:healpix_disc}

Although recent work favours a disc origin for these structures, the debate over its origin (the alternative being that these are the tidal debris of an accreted MW satellite) is still on going. Based solely on the morphology that our method allows to observe, both MNC and ACS could very well be different wraps of the same tidal stream. 
If that was the case then we should be able to see at least a hint of continuity in the south, unless the tails only emerge from behind the disc at Galactic longitudes where the disc is already too dense for our method to detect them. \citet{Penarrubia2005} presented an N-body model fitted to the observations of MNC available at the time, which was later used by \citet{Slater2014} to show that there is a broad agreement with the PanSTARRS-1 \citep{Chambers2016} data. We find that the arch described by the debris generated with their model is too wide to explain MNC as we detected it\footnote{Yet, the shape that we recover could be affected by the \Gaia scanning law, as we mentioned in \ref{sec:mnc&acs}} but we note the presence of a tail (top right panel of Fig.~5 from \citealt{Slater2014}) that resembles ACS. Comparing the morphology now with the N-body simulation of \citet{Laporte2019}\footnote{We tried to compare also with the simulations of \citet[][Fig.~5]{Kazantzidis2008} and \citet[][Fig.~4]{Gomez2016} but there are too few particles to make a good assessment given the level of detail that the data is providing.}, where these "feathers" appear as a result of the interaction with Sgr, we note that the overall agreement is good (top left panel of their Fig.~1). For instance, the difference in latitude between the two structures and the fact that the one on top stops abruptly at a given longitude. Nonetheless, the "feather" corresponding to ACS obtained with their N-body simulations is thicker than the observed one and does not present a higher density of stars close to the turning point of the vertical oscillation (the highest point in latitude) as we see in the data. However, this could simply be due to resolution limitations and the fact that these simulations were not meant to be an exact match to the MW. 

\begin{figure}[h!]
    \centering
    \includegraphics[width=\linewidth]{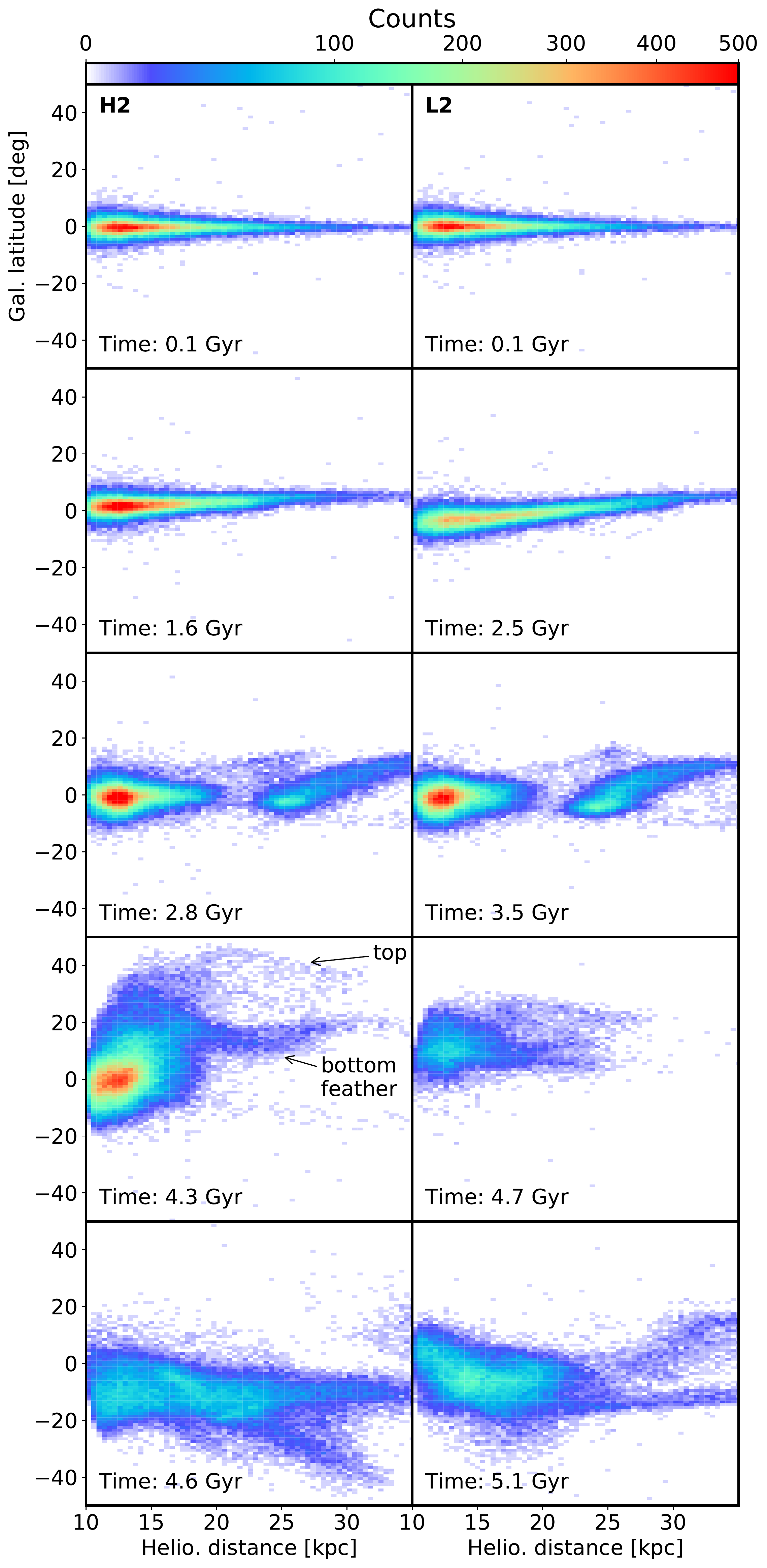}
    \caption{Heliocentric distance against Galactic latitude for different snapshots of the simulations by \citet{Laporte2018}. On the left, the H2 simulation and, on the right, the L2. In both cases, the Sun has been placed at (x=-8\,kpc, y=0\,kpc). The stars shown in each panel correspond to those in slices of Galactic longitude 120$^\circ$ < l < 240$^\circ$. The snapshot at 4.3\,Gyr for H2 corresponds to the one where ACS- and MNC-like structures (top and bottom feathers, respectively) have been reported in \citet{Laporte2019}.}
    \label{fig:laporte_sim}
\end{figure}

To explore a little bit deeper the simulations of \citet{Laporte2018}, in Fig.~\ref{fig:laporte_sim} we show the plane of Galactic latitude against Heliocentric distance of the particles that are at Galactocentric radius >\,18\,kpc. To do that, we locate the Sun at (x,y)=(-8, 0)\,kpc and look at the particles with 120$^\circ$ < $\ell$ < 240$^\circ$. The snapshots are chosen such that we can see the outer disc at the beginning of the simulation (first row), right after the first pericentric passage of Sgr (second), at the time of first apocentre (third), right after the second pericentre (fourth) and, finally, at the second apocentre (fifth). The first thing we note is that, while the first pericentre passage seems to have a similar effect in both simulations, causing some material to be ejected from the system (third row), the difference becomes more noticeable with the second passage. In particular, we can see how in the H2 simulation a couple of "feathers" appear in the fourth row, with roughly constant declination ($\sim$20$^\circ$ and $\sim$40$^\circ$) and a large extension in distance. These correspond to the MNC- and ACS-like structures reported in \citet{Laporte2019}. The shape that we observe for MNC in the \Gaia data seems more concentrated in distance while being more extended in latitude (see Fig.~\ref{fig:bVSgmag_130-150}). Nevertheless, a more detailed analysis of the distances in our data, as mentioned in previous sections, is still necessary to produce a more precise 3D characterisation of MNC and ACS. 

Secondly, we note that, even at the timescales of $\sim$100\,Myr, the large-scale distribution of stars in the whole outer disc of the simulated MW changes significantly with time in both the radial and vertical directions. In turn, this would suggest that the phase-space configuration of the outer disc and of its feathers at present time can in principle pose strong constraints on the time-evolution of the perturbation (assuming that they were caused by a single perturber). The data that we have obtained in this work, being precise, continuous and covering a large range of longitudes and latitudes, can be used for describing MNC and ACS with analytical/semi-analytical models like, for instance, with the analytic method presented in \citet{Weinberg1998}. This efficient way of exploring the parameter space could be used to obtain a quantitative measure of the goodness of fit for each model, beyond the small set of simulations analysed here. Such an exercise is crucial as it would allow us to predict where the continuation of these structures should appear, both in Galactic coordinates as well as in distance, providing a way to search for them actively. More importantly, we could quantify the mass of the perturber and the history of its orbit. Or even, as mentioned in \citet{Laporte2019}, use MNC and ACS to constrain the Galactic potential: the rotation curve at that distance, its slope, the shape of the Dark Matter halo, etc.

Another option to compare models objectively is to add particles generated with N-body models, one that could reproduce the observed MNC and ACS, to a mock catalogue of the Galaxy. In this work we have used the \citet{Rybizki2018} catalogue as an example of Galaxy without substructure, but we now know that it was not so representative of the MW and, also, that it underestimated the observational errors \citep{Rybizki2020}. The recent  \citet{Rybizki2020} catalogue has fixed much of these issues and provides a mock to use with the up-coming \Gaia EDR3 \citep{Brown2019}. With these approach, and taking into account the nuances of our methodology, we can attempt a quantitative comparison with the models. One of the key parameters to generate the N-body models would be the stellar mass contained inside these structures, which is currently poorly measured \citep{Morganson2016}. In this work, we have attempted to quantify the fraction of the disc that is within MNC and ACS in Fig.~\ref{fig:bVSratio}, but it is just a rough estimate based solely on the number of giants.

We have also studied the kinematic information obtained with the proper motion of the peaks, and we see that MNC and ACS rotate slightly slower than the disc at the solar position, in agreement with \citet{deBoer2018}. Nevertheless, this alone does not prove that these structures were once part of the disc. Since the two mechanisms proposed (extra-galactic/internal) have distinct formation time-scales and chemical properties, by exploring also their CMDs we can more easily distinguish between both. In this sense, the ratio of RR Lyrae to M giants that we find (<5\%) is unlikely for structures composed of tidal debris from an accreted satellite. This adds to the recent studies of \citet{Bergemann2018} and \citet{Laporte2020}, where it is shown that the abundances, the distribution in the [Mg/Fe] vs [Fe/H] plane and the mean metallicity of MNC and ACS are inconsistent with the extragalatic scenario.

If indeed MNC and ACS were once part of the disc, then we can now use them as chemical fossils, an idea already exposed in \citet{Laporte2020}. After they were kicked out of the disc, the stellar formation of these structures most likely came to a halt as any gas that initially accompanied the stars must have quickly settled back to the disc thanks to its efficient energy dissipation mechanisms. As a result, their current population is a frozen relic of the outskirts of the MW at the time when the perturbation occurred. With enough spectroscopic abundances we could learn about the gas that dwelled at the edge of our Galaxy some Gigayears ago and use that information to constrain the chemo-dynamical models of the MW. 


\section{Conclusions}\label{sec:healpix_conc}

The application of the WT to the proper motion space has proven extremely useful to reveal the kinematic substructure of the halo and outer disc. By removing most of the foreground with a simple yet effective cut in parallax, our method is able to efficiently detect kinematic substructure in the halo and even external galaxies like M33 or the Magellanic clouds, several dwarf spheroidals and dozens of globular clusters, as well as the Sgr stream. It has also revealed the sharpest picture of the anticentre, with MNC and ACS appearing as the third most prominent structures in the distant sky (only after the Magellanic clouds and Sgr). 

We have been able to blindly detect the whole MNC north as well as the ACS from l\,$\sim$\,120$^\circ$ to l\,$\sim$\,230$^\circ$. Our findings are in good agreement with previous studies like \citet{Laporte2020}, who also used DR2 data to investigate these structures. Nevertheless, we have been able to characterise their morphology with great detail, which is crucial for obtaining the orbital parameters of these groups of stars. We observe MNC with an arch-like shape, broader at small longitudes and becoming thinner towards larger longitudes. Nevertheless, the RC stars that we have selected can be seen to span a wider range of latitudes, therefore a detailed study of the selection function of \Gaia and the extinction is needed to confirm how much of this shape is cause by the scanning law. ACS can be seen at larger latitudes than MNC throughout the whole longitude range where we detect them, and has a maximum of relative intensity when it reaches the highest latitude at l$\sim$140$^\circ$ (consistent with a pile up of stars at the maximum height in the orbit), and stops abruptly at a longitudes of $\sim$110$^\circ$. This behaviour, added to the fact that we do not observe a clear continuity in the south, favours the perturbative scenario proposed by \citet{Ibata2003}, later supported by the simulations of many other authors \citep[e.g., ][]{Gomez2016,Laporte2019}. Moreover, the kinematics of these features, which differ from the bulk motion of the disc stars that lay in front, are compatible with a low eccentricity orbit at $\sim$10\,kpc that rotates similarly to the disc. 

By analysing the apparent magnitude of the RC stars selected by proper motion we have been able to trace MNC down to a latitude of $\sim$5$^\circ$, closer to the disc than ever before. Also, by measuring the median apparent magnitude of the RC stars of each structure and converting to Heliocentric distance, we have determined that ACS ($\sim$11.7\,kpc) is roughly 1\,kpc farther away from the Sun than MNC ($\sim$10.6\,kpc). This actually means that both structures are at roughly the same Galactocentric radius (but at heights above the disc of, respectively, $\sim$6.5\,kpc and $\sim$4.5\,kpc). Also, we have shown that MNC and ACS, despite being different structures, are extended in distance and in the sky, and their tails overlap both in the 3D physical space as well as in kinematic space.

In the south, we have found a diffuse population of giants at 130$^\circ$ < l < 150$^\circ$ and 190$^\circ$ < l < 210$^\circ$, coinciding with the regions of low extinction, that we do not observe in the mock catalogue nor in the north. Their apparent magnitudes span the range 14\,<\,$G$\,<\,15.5\,mag which implies distances for a RC star not affected by extinction between 5 and 10\,kpc. These could be related with the vertical wave described in \citet{Xu2015}, and are most likely the so-called MNC south first reported by \citet{Ibata2003}. On the other hand, in the longitude range 130$^\circ$ < l < 150$^\circ$ we have observed a faint trace of RC at $G\sim$17\,mag (heliocentric distance $\sim$16\,kpc) that most likely corresponds to the TriAnd overdensity. Nevertheless, due to the contamination of nearby stars close to the disc and large the distance uncertainties, we have not been able to explore the morphological connection between these structure and MNC north.

Studies like this will benefit the most from the next \Gaia release (EDR3) which is expected to contain proper motions twice as precise (on average), and increase the number of stars at the faintest magnitudes \citep{Brown2019}. As a result, the structures that we detect will become more concentrated in the proper motion space, and in turn produce stronger signals in our maps of relative intensity. Also, the effects of the scanning law should diminish as a result of the additional year of observations. With it, we should be able to remove the foreground contamination more efficiently and detect the anticentre structures continuously at all latitudes, providing a direct observation of their 3D morphology. Moreover, the WEAVE spectroscopic survey will also observe this region, so we could potentially obtain radial velocities and abundances for a large fraction of the giants in our sample. This means that we will be able to trace MNC and ACS, probably even TriAnd, more clearly and deeper, and obtain a less contaminated sample of members.

The challenge now is to find the way to use our data to form a coherent and unified picture of the outer disc, constraining the properties of the different agents involved (MW disc, dark matter halo, Sgr), as well as to prepare these methods to work with the future samples that will soon become available.

\begin{acknowledgements}
    This work has made use of data from the European Space Agency (ESA) mission {\it Gaia} (\url{https://www.cosmos.esa.int/gaia}), processed by the {\it Gaia} Data Processing and Analysis Consortium (DPAC, \url{https://www.cosmos.esa.int/web/gaia/dpac/consortium}). Funding for the DPAC has been provided by national institutions, in particular the institutions participating in the {\it Gaia} Multilateral Agreement. 
    This work has been supported by the Agence Nationale de la Recherche (ANR project SEGAL ANR-19-CE31-0017). It has also received funding from the project ANR-18-CE31-0006 and from the European Research Council (ERC grant agreement No. 834148).
    This project has received funding from the European Union's Horizon 2020 research and innovation programme under the Marie Sk{\l}odowska-Curie grant agreement No. 745617 and No. 800502.
    This work was supported by the MINECO (Spanish Ministry of Economy) through grant ESP2016-80079-C2-1-R and RTI2018-095076-B-C21 (MINECO/FEDER, UE), and MDM-2014-0369 of ICCUB (Unidad de Excelencia 'Mar\'\i a de Maeztu').
     This project has received support from the DGAPA/UNAM PAPIIT program grant IG100319.
    CM is grateful for the hospitality of the ICCUB, during visits in which part of this research was carried out.  
\end{acknowledgements}

%
%

\bibliographystyle{aa}
\bibliography{mybib}

\begin{appendix}

\section{Queries to the Gaia archive}\label{app:queries}

To obtain the mean quantities used in this work all around the sky and in a single file, we use the following query to the \Gaia Archive:

\begin{lstlisting}
SELECT gaia_healpix_index(5, source_id) AS healpix_5, count(*) as N, avg(astrometric_n_good_obs_al) AS avg_n_good_al, avg(astrometric_gof_al) AS avg_gof_al, avg(astrometric_excess_noise) as, avg_excess_noise, avg(bp_rp) as avg_bprp, avg(phot_g_mean_mag) as avg_g, avg(pmra) as avg_pmra, avg(pmdec) as avg_pmdec, avg(pmra_error) as avg_pmra_error, avg(pmdec_error) as avg_pmdec_error
FROM gaiadr2.gaia_source WHERE parallax-parallax_error < 0.1 AND BP_RP>0.2 GROUP BY healpix_5 
\end{lstlisting}

\noindent Computing the standard deviations is not so straight forward, though, as there is no implemented function to do so in ADQL. Instead, we repeat the same query but this time using the average \emph{squared} of the quantities. A simple subtraction of both tables returns the standard deviations:

\begin{equation}
    \sigma = \sqrt{E[X^2] - (E[X])^2}
\end{equation}
\noindent where $E[X]$ is the mean of a vector $X$.

\begin{figure}[h!]
   \centering
    \includegraphics[width=0.95\linewidth]{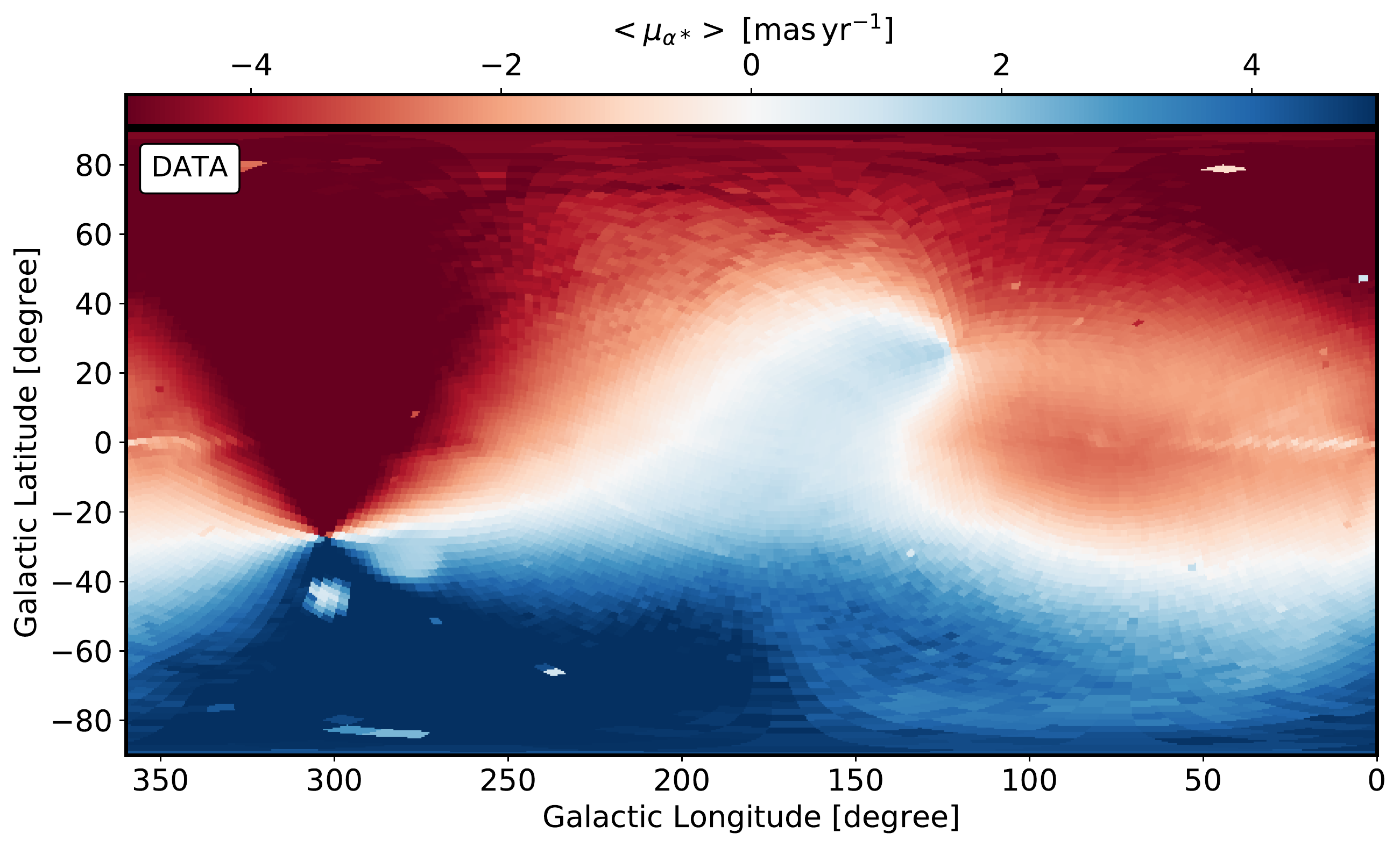}
    \includegraphics[width=0.95\linewidth]{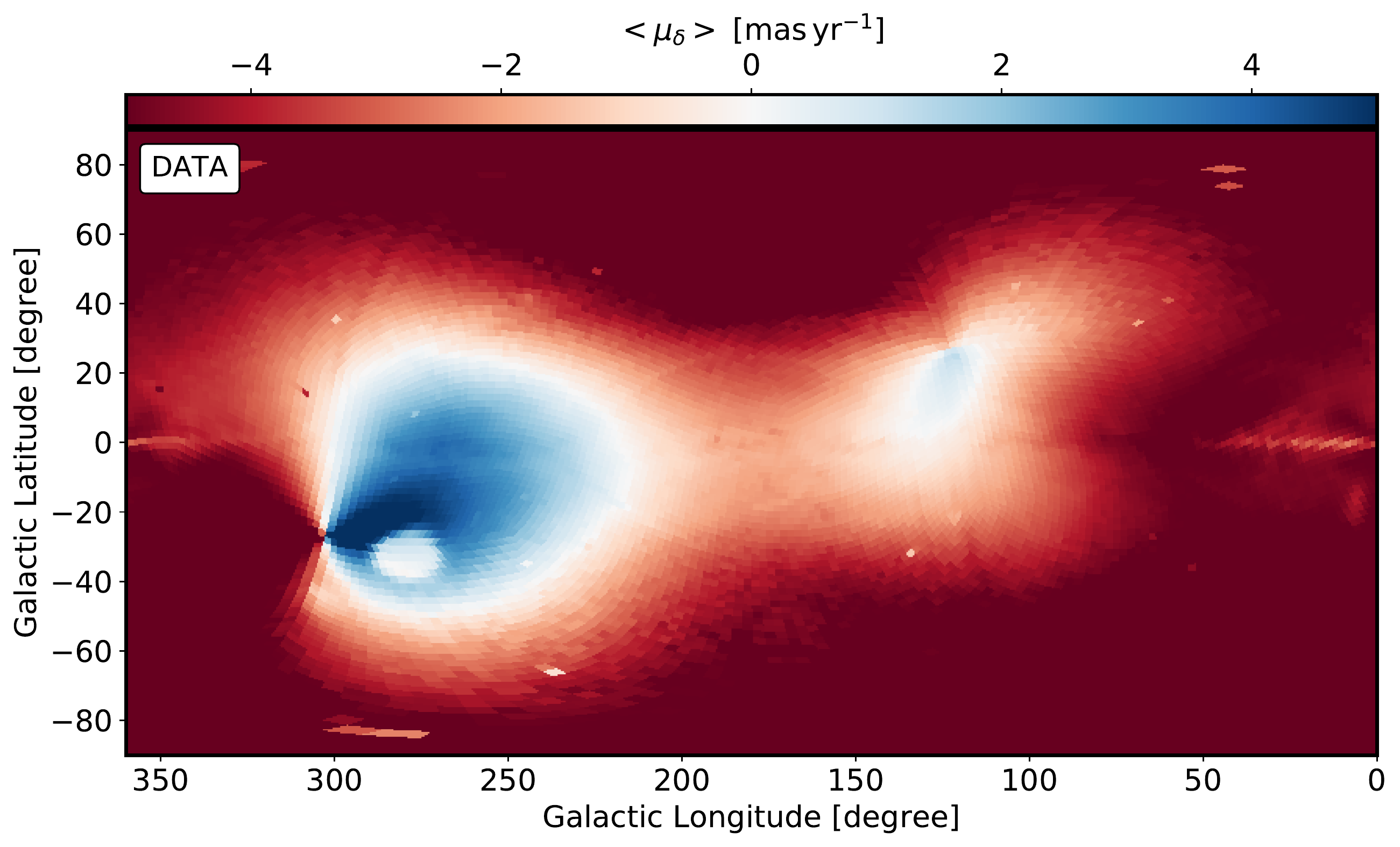}
    \caption{Average proper motion of the stars selected according to the criteria set in Sect.~\ref{sec:data&methods} as a function of position in the sky. Top: Proper motion in right ascension. Bottom: Same but in declination. Only a few globular clusters, the Large Magellanic cloud and the most inner parts of the Sgr core are noticeable, as the rest of the field is dominated by the solar reflex (peculiar motion of the Sun and rotation of the Local Standard of Rest).}
    \label{fig:allsky-avg_pm} 
\end{figure}

\begin{figure}[h!]
   \centering
    \includegraphics[width=0.95\linewidth]{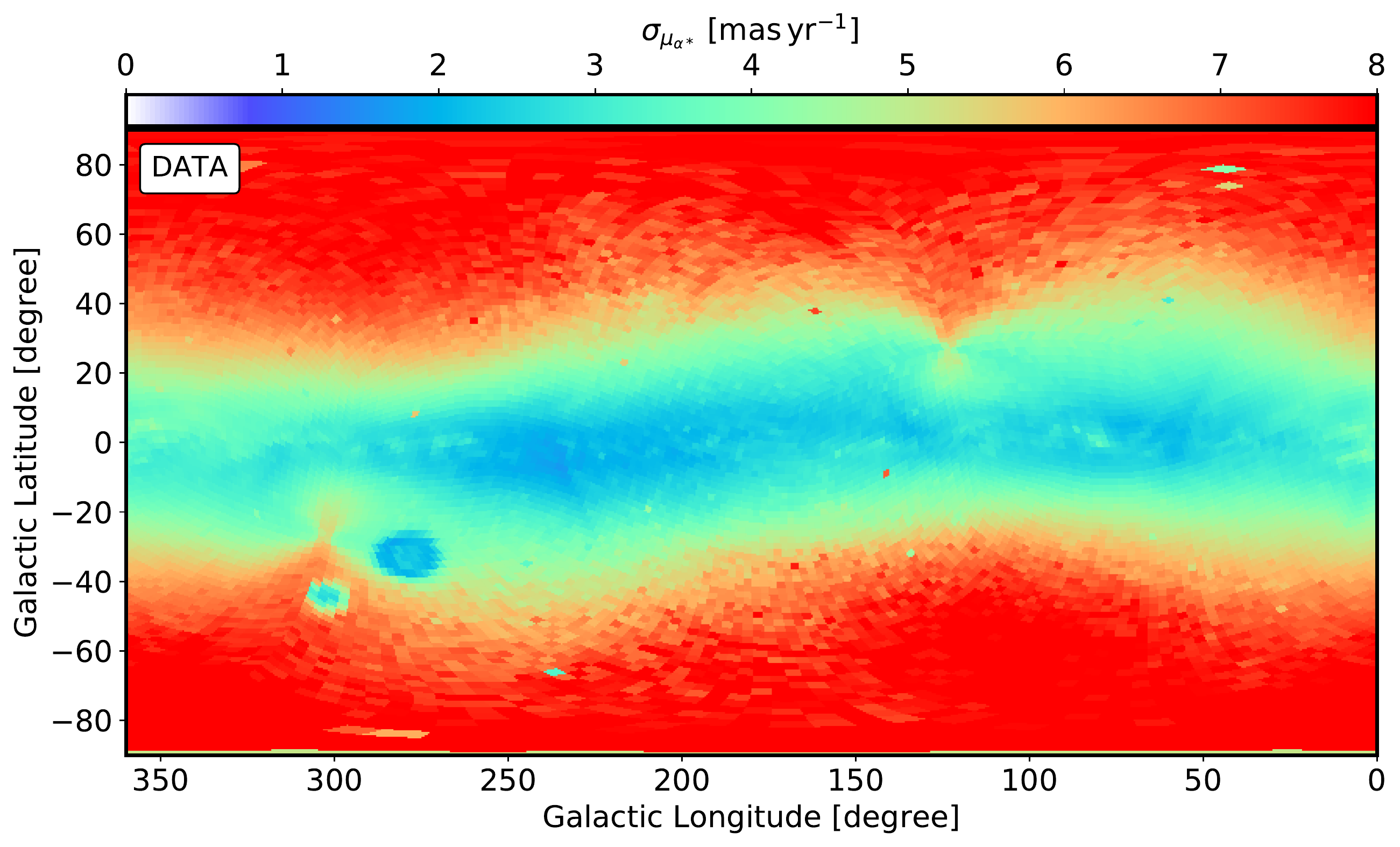}
    \includegraphics[width=0.95\linewidth]{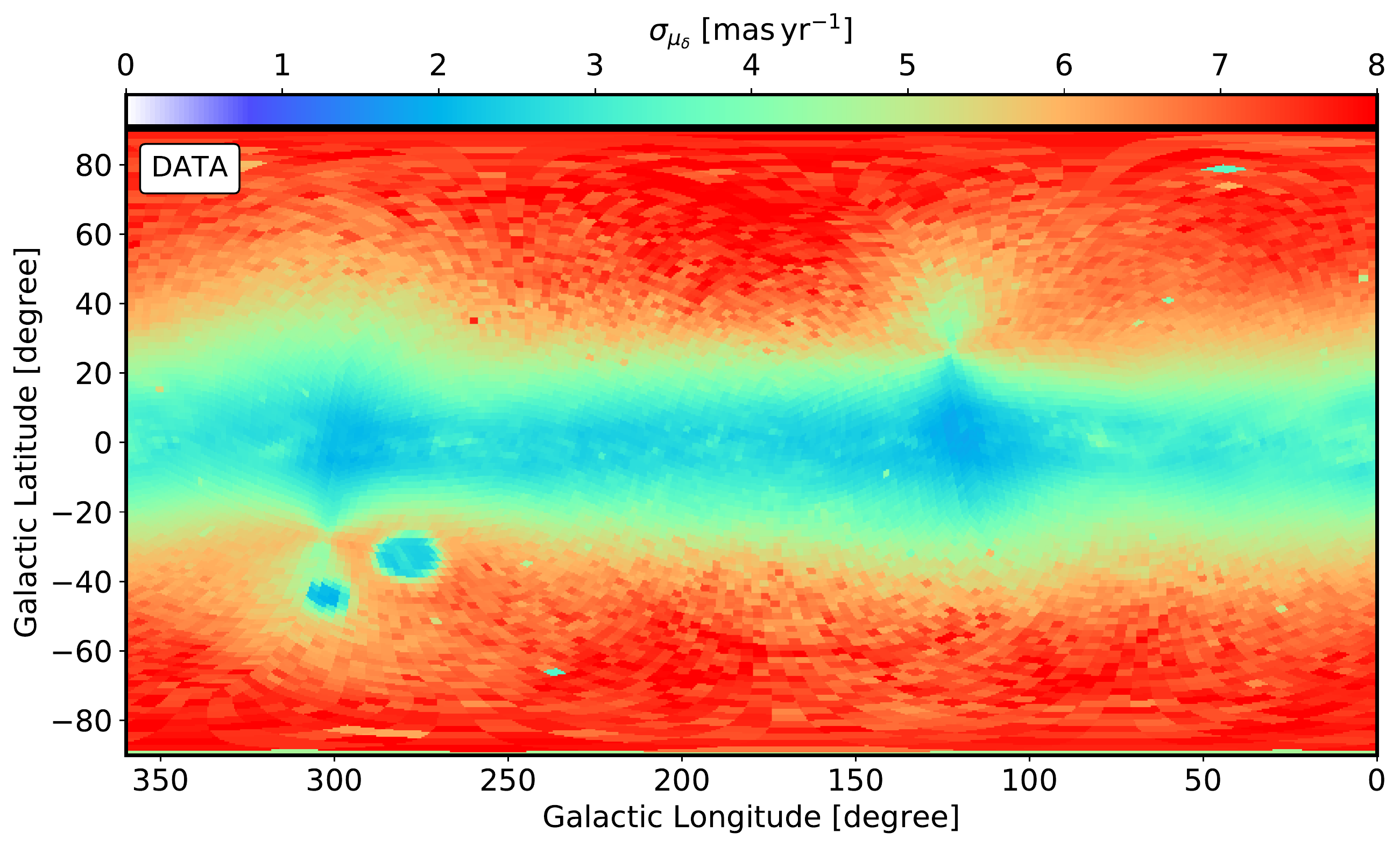}
    \caption{Dispersion in proper motion of the stars selected according to the criteria set in Sect.~\ref{sec:data&methods} as a function of position in the sky. Top: Proper motion in right ascension. Bottom: Same but in declination.}
    \label{fig:allsky-std_pm} 
\end{figure}

Figures \ref{fig:allsky-avg_pm} and \ref{fig:allsky-std_pm} contain, respectively, the average proper motion and its dispersion for both components, right ascension and declination. With the mean proper motions we basically see a kinematic field dominated by the combination of the solar peculiar motion and the rotation of the Galaxy. Given the large amount of nearby stars that our filter (Eq.~\ref{eq:parallax_filter}) cannot remove, only the very dense structures can be seen here: the Magellanic clouds, some globular clusters and, although very faint, the Sgr stream. To understand better the extent of the disc in our sample, we use the relation between the kinematic dispersion and the kinetic temperature of each population (disc vs halo). We note a transition at $\sim$30$^\circ$ in latitude from a rather cold population (disc) to a hotter one (halo). 

\section{Gaia DR2 mock catalogue}\label{app:mock}

To test the signal that we would expect to see in a galaxy without substructure, we run our whole method to the proper motion histograms obtained from a mock catalogue. For that, we query the \citet{Rybizki2018} catalogue and download, for each \HPf, up to 2\,000\,000 stars. Among the quantities available to download, we select all the astrometry and photometry, as well as the age of the stars which we then use to separate the stars in thin disc, thick disc and halo. Then, we draw for each star one realisation from a normal distribution centred on the true values and with a dispersion equal to the provided observational uncertainties to produce the mock particles. In the case of the colours, we apply the errors to the fluxes and then convert the observed fluxes to magnitudes using the equations published in the \Gaia web-page\footnote{\href{https://www.cosmos.esa.int/web/gaia/dr2-known-issues}{https://www.cosmos.esa.int/web/gaia/dr2-known-issues}} \citep{MAW2018}. We note that these are meant to be used with synthetic fluxes derived using the same pass-bands, which is not the case here, but since we only use the photometry for reference (the cut in colour has little impact), we do not need a perfect match with reality.

Once we have the mock particles, we apply the cuts in parallax and colour described in Sect.~\ref{sec:data&methods}, generate the proper motion histograms and analyse them with the WT in the same manner as we did for the data. The value of the WT coefficient is sensitive to the absolute number of counts and, in consequence, we scale the histograms such that the sum of all the bins equals the number of stars observed in that same \HP with \Gaiaf. No substructure is added in this way since a scaling of the histogram does not bias the centroid and, whenever we show the coefficients we do it after normalising by the number of stars in the \HPf.

\begin{figure}
   \centering
  \includegraphics[width=.5\textwidth]{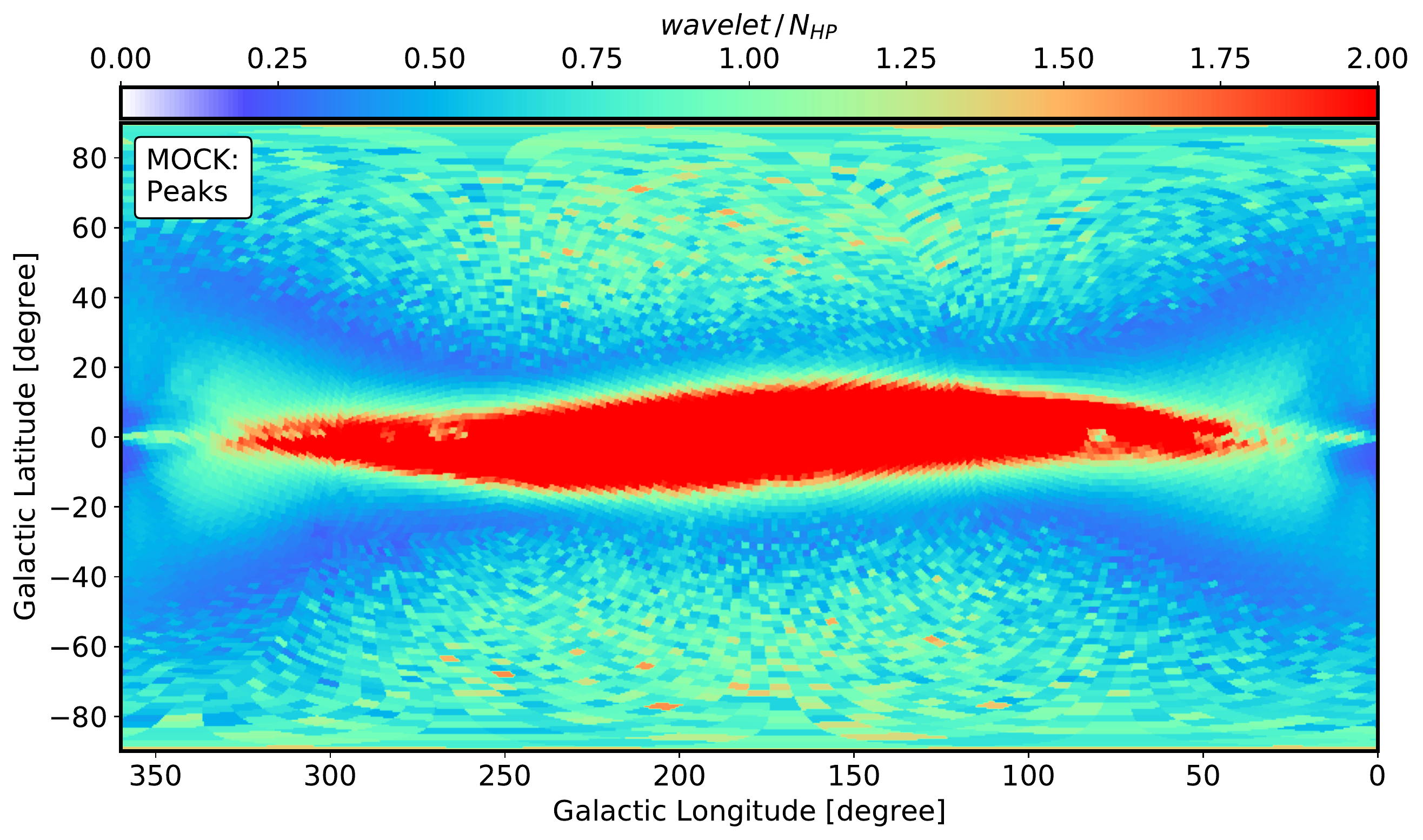}
  \caption{Relative intensity of the dominant structure in the proper motion plane of each \HP for the mock catalogue.}
     \label{fig:allsky-wt_mock} 
\end{figure}

\begin{figure}
   \centering
    \includegraphics[width=0.5\textwidth]{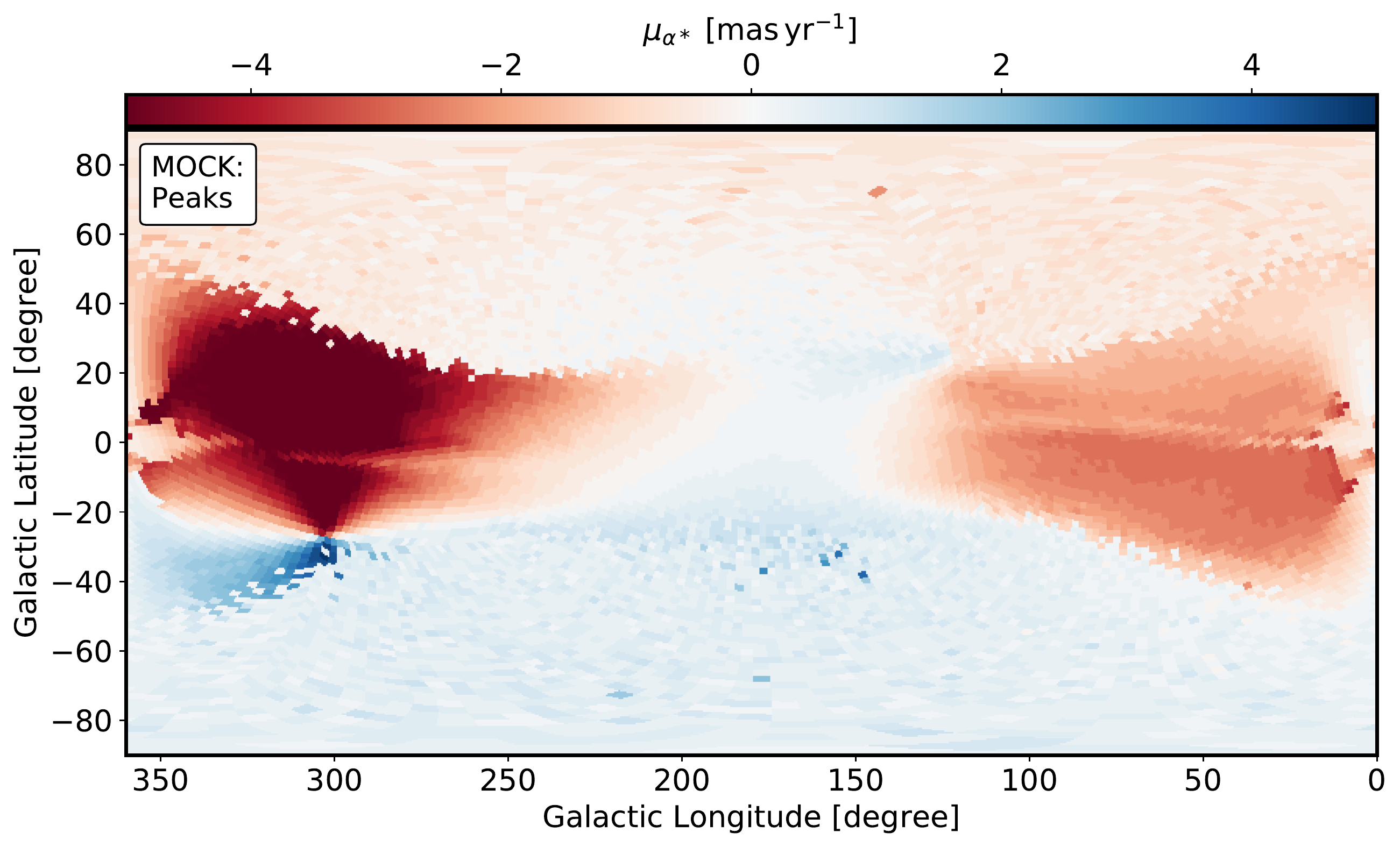}
    \includegraphics[width=0.5\textwidth]{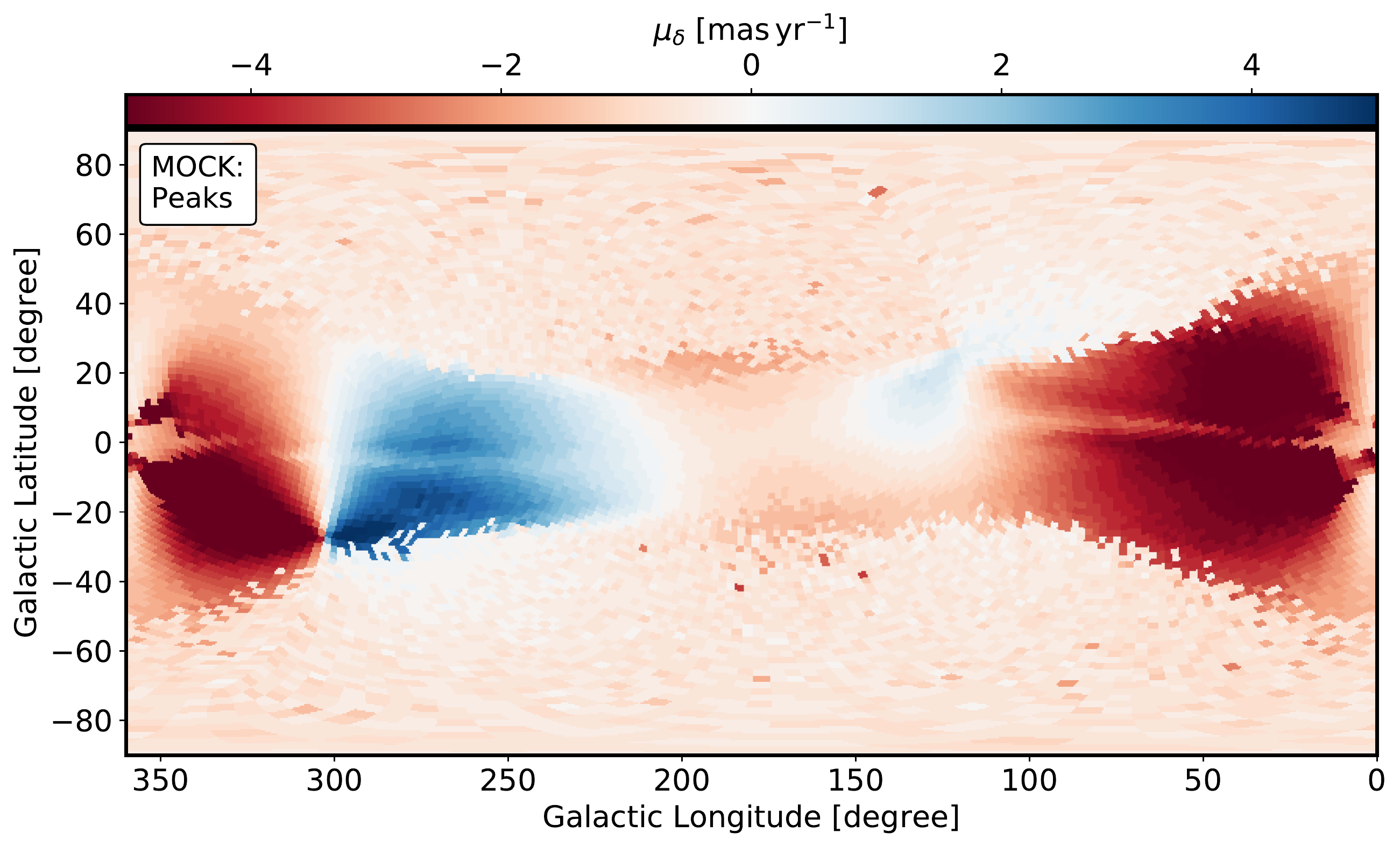}
    \caption{Same as Fig.~\ref{fig:allsky-pm} but for the mock.}
    \label{fig:allsky-pm_mock} 
\end{figure}

Figure~\ref{fig:allsky-wt_mock} shows the result of applying our methodology to the mock catalogue. The only structure present is the geometrical warp introduced in the underlying model of the galaxy. Apart from that, we note a sharp transition between the disc and the halo, noticeable as a drastic change in the relative intensity. There is also a change between the thin disc that dominates the anticentre, and the thick disc that dominates the central parts of the mock MW. If we analyse the proper motions of the peaks obtained (Fig.~\ref{fig:allsky-pm_mock}) we recover the reflex of the solar motion, with the location of the poles of the equatorial sphere clearly visible as singular points, and the perspective effect caused by the rotation, which introduces a gradient in the proper motions with Galatic longitude and latitude. This is the reason why the transition to the halo is so sharp: once the dominant structure in the proper motion plane is the halo, we observe the reflex of a non-rotating stellar system which only has 4 lobes instead of 8 as in the case of the disc.

\begin{figure*}
    \centering
    \includegraphics[width=\linewidth]{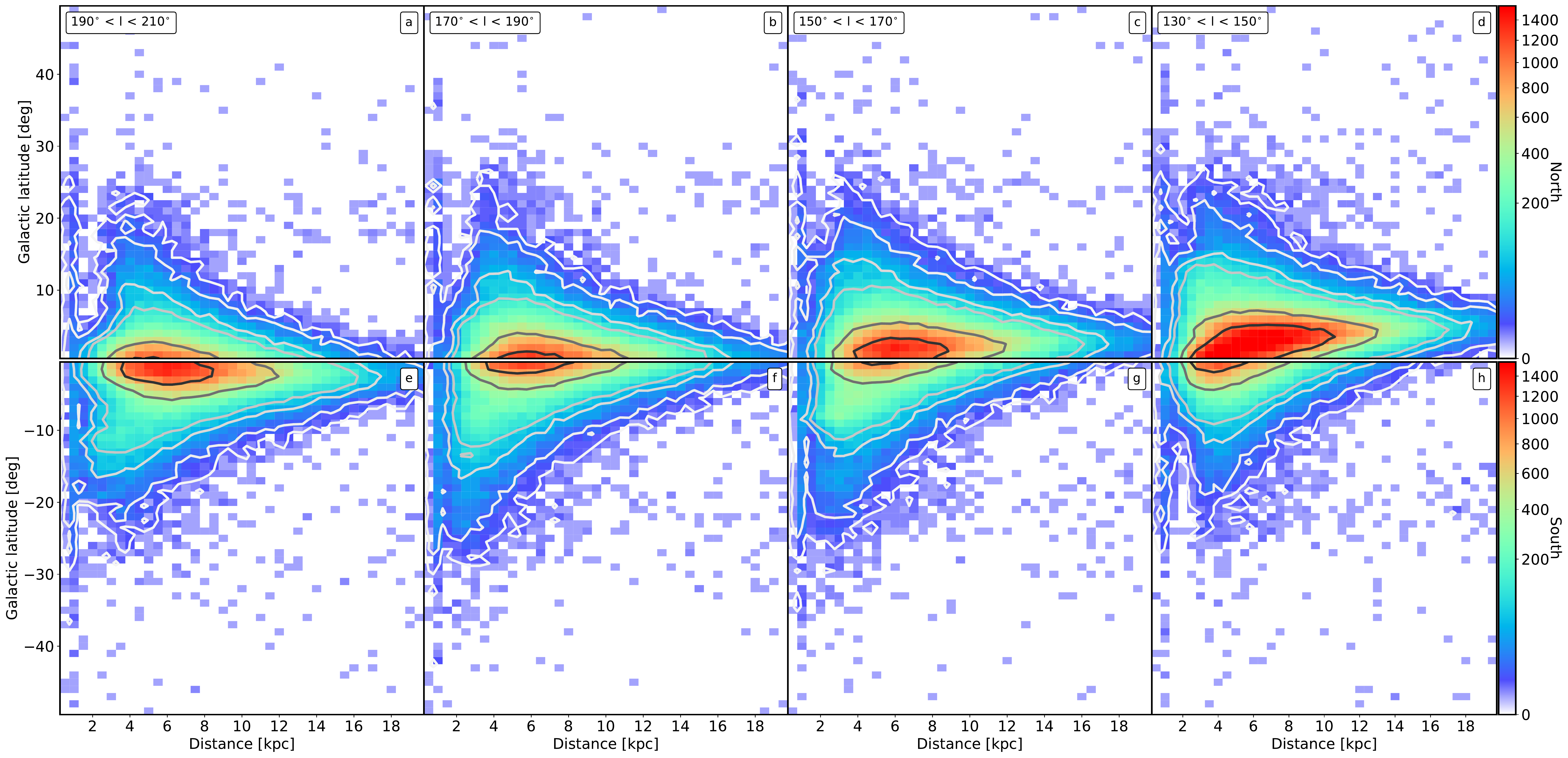}
    \caption{Same as Fig.~\ref{fig:bVSdist_data} but for the particles in the mock catalogue that fall inside the peaks detected in the data.}
    \label{fig:bVSdist_mock}
\end{figure*}

In some of the figures, e.g. Figs~\ref{fig:patches_comparison} or \ref{fig:bVSdist_mock}, we select the particles not according to the peak obtained with the mock but with coordinates of the peaks detected in the data. In doing so we can check what is distribution in the CMD or in distance of the particles that have the observed kinematics.

\end{appendix}

\end{document}